{}
{}

\def\verdraft{3}
\def\verpublish{4}
\def\verarxiv{5}

\def\version{5}

\documentclass[12pt,a4paper]{article}

\usepackage{amsmath,amssymb}
\ifnum\version<2\usepackage{showkeys}\fi
\usepackage{graphicx}
\usepackage[bookmarks=true,hyperfigures=true]{hyperref}
\usepackage[usenames,dvipsnames]{color}
\usepackage[nosort]{cite}
\usepackage[bulletsep]{collref}
\usepackage[utf8]{inputenc}

\def\showkeyslabelformat#1{{\usefont\encodingdefault\familydefault%
 \seriesdefault\shapedefault\normalfont\tiny\ttfamily#1}}
\def\showkeysrefformat#1{{\normalfont\tiny\ttfamily#1}}
\makeatletter
\def\SK@@ref#1>#2\SK@{%
 {\@inlabelfalse\leavevmode\vbox to\z@{%
 \vss\SK@refcolor\rlap{\vrule\raise .75em%
  \hbox{\showkeysrefformat{#2}}}}}}
\makeatother

\usepackage[a4paper,text={160mm,247mm},centering]{geometry}

\RequirePackage{verbatim}

\makeatletter
\ifnum\version<\verarxiv\relax
\RequirePackage{shellesc}
\ifnum\ShellEscapeStatus>0\relax
\AddToHook{enddocument/afteraux}{\ShellEscape{bibtex \jobname}}
\fi
\newwrite\bibinl@out
\newenvironment{bibtex}[1][\jobname]{%
  \immediate\openout\bibinl@out #1.bib
  \immediate\write\bibinl@out{\@percentchar generated from `\jobname' starting line \the\inputlineno^^J}%
  \def\verbatim@processline{\immediate\write\bibinl@out{\the\verbatim@line}}%
  \@bsphack\let\do\@makeother\dospecials\catcode`\^^M\active\verbatim@start
}%
{\immediate\closeout\bibinl@out\@esphack}
\else
\newenvironment{bibtex}[1][\jobname]{\comment}{\endcomment}
\fi
\makeatother

\usepackage{eqnlines}
\eqnlinesset{number=on,numberline=center}
\eqnlinesset{punctall={,,.}}

\counterwithin{equation}{section}

\usepackage[font=small,labelfont=bf,width=0.85\textwidth]{caption}

\ifnum\version<\verarxiv\else\PassOptionsToPackage{write=false,compile=false}{mpostinl}\fi
\ifnum\version=\verarxiv\else\PassOptionsToPackage{write=true,compile=true}{mpostinl}\fi
\ifnum\version<\verpublish\else\PassOptionsToPackage{labelnames}{mpostinl}\fi
\ifnum\version>\verdraft\else\PassOptionsToPackage{warnunused=true}{mpostinl}\fi
\ifnum\version<1\PassOptionsToPackage{now=true,nowall=true}{mpostinl}\fi
\PassOptionsToPackage{clean=false}{mpostinl}

\RequirePackage{graphbox}
\RequirePackage[clean=false]{mpostinl}

\begin{mposttex}
\usepackage{amsmath,amssymb}
\usepackage[utf8]{inputenc}
\usepackage[usenames,dvipsnames]{color}
\end{mposttex}

\begin{mpostdef}
def pensize(expr s)=withpen pencircle scaled s enddef;
def fillshape(expr p,c)=
  fill p withcolor c;
  draw p
enddef;
def drawdot(expr z,s)=
  fill fullcircle scaled s shifted z
enddef;
def filldot(expr z,s,ci)=
  fillshape(fullcircle scaled s shifted z, ci)
enddef;
def drawcross(expr z,s,r,t,c)=
  draw ((-0.5,-0.5)--(+0.5,+0.5)) scaled s rotated r shifted z pensize(t) withcolor c;
  draw ((+0.5,-0.5)--(-0.5,+0.5)) scaled s rotated r shifted z pensize(t) withcolor c;
enddef;
\end{mpostdef}

\begin{mpostdef}
pair vpos[];
numeric nums[];
path paths[];
newinternal numeric xu,yu;
xu:=1cm;
ahlength:=7pt;
\end{mpostdef}

\begin{mpostdef}
def midarrowperc (expr p, t) =
  fill arrowhead subpath(0,arctime(t*arclength(p)+0.5ahlength) of p) of p
enddef;
def midarrow (expr p, t) =
  fill arrowhead subpath(0,arctime(arclength(subpath (0,t) of p)+0.5ahlength) of p) of p
enddef;
\end{mpostdef}

\begin{mpostdef}
color colone,coltwo;
colone:=0.7[green,black];
coltwo:=0.6[blue,black];
vardef drawone expr p =
  draw p pensize(1.25pt) withcolor colone;
enddef;
vardef drawtwo expr p =
  draw p pensize(1.75pt) withcolor coltwo;
enddef;
def arrowone (expr p, t) =
  midarrow(p,t) withcolor colone;
enddef;
def arrowtwo (expr p, t) =
  midarrow(p,t) withcolor coltwo;
enddef;
\end{mpostdef}

\begin{mpostdef}
def lambdapath(suffix i)(expr a,b,s)(text f)=
for i:=a step s until (b+0.5s): if i<>a: .. fi f endfor
enddef;
\end{mpostdef}

\ifnum\version=11
\begin{mposttex}[dual]
\usepackage[bitstream-charter,cal=cmcal]{mathdesign}
\end{mposttex}
\fi

\begin{mposttex}[dual]

\usepackage{mathfixs}
\ProvideMathFix{autobold,greekcaps,frac,root}
\ProvideMathFix{multskip,diff,der,defeq,eqdef}
\ProvideMathFix{numset,numsets}
\ProvideMathFix{vfrac,rfrac,iunit*nb}
\ProvideMathFix{reim,tr,diag}

\RequirePackage[extdef]{delimset}

\newcommand{\half}{\rfrac{1}{2}}
\newcommand{\ihalf}{\rfrac{\iunit}{2}}

\newcommand{\alg}[1]{\mathfrak{#1}}
\newcommand{\gen}[1]{\mathrm{#1}{}}
\newcommand{\yanghat}[1]{\widehat{#1}}
\newcommand{\genyang}[1]{\yanghat{\gen{#1}}{}}
\newcommand{\local}{\text{loc}}
\newcommand{\bilocal}{\text{biloc}}

\newcommand{\superN}{\mathcal{N}}
\newcommand{\oper}[1]{\mathcal{#1}}

\newcommand{\action}{\oper{S}}
\newcommand{\var}{\delta}
\newcommand{\nfield}[1]{{[#1]}}

\end{mposttex}

\providecommand{\href}[2]{#2}
\newcommand{\arxivlink}[1]{\href{http://arxiv.org/abs/#1}{arxiv:#1}}

\makeatletter
\def\mr@ignsp#1 {\ifx\:#1\@empty\else #1\expandafter\mr@ignsp\fi}%
\newcommand{\multiref}[1]{\begingroup
\xdef\mr@no@sparg{\expandafter\mr@ignsp#1 \: }%
\def\mr@comma{}%
\@for\mr@refs:=\mr@no@sparg\do{\mr@comma\def\mr@comma{,}\ref{\mr@refs}}%
\endgroup}
\renewcommand{\eqref}[1]{(\multiref{#1})}
\makeatother

\makeatletter
\newcommand{\namedref}[2]{\hyperref[#2]{#1~\ref*{#2}}}
\newcommand{\secref}{\@ifstar{\namedref{Section}}{\namedref{Sec.}}}
\newcommand{\appref}{\@ifstar{\namedref{Appendix}}{\namedref{App.}}}
\newcommand{\tabref}{\@ifstar{\namedref{Table}}{\namedref{Tab.}}}
\newcommand{\figref}{\@ifstar{\namedref{Figure}}{\namedref{Fig.}}}
\makeatother

\let\oldbib=\thebibliography
\def\thebibliography{\phantomsection\addcontentsline{toc}{section}{\refname}\oldbib}

\let\oldtoc=\tableofcontents
\def\tableofcontents{\phantomsection\addcontentsline{toc}{section}{\contentsname}\oldtoc}

\providecommand{\hypersetup}[1]{}

\hypersetup{plainpages=false}
\hypersetup{pdfpagemode=UseNone}
\hypersetup{bookmarksnumbered=true}
\hypersetup{pdfstartview=FitH}
\hypersetup{colorlinks=false}
\hypersetup{citebordercolor={.5 1 .5}}
\hypersetup{urlbordercolor={.5 1 1}}
\hypersetup{linkbordercolor={1 .7 .7}}

\ifnum\version<\verpublish\PassOptionsToPackage{draft}{metastr}\fi
\RequirePackage{metastr}

\ifnum\version<\verdraft\metasetterm{draft}{EDITING}\fi

\usepackage{ifpdf}
\ifpdf\else\RequirePackage[active]{srcltx}\fi
\RequirePackage{color}

\newcommand{\remark}[2][]{}

\newcommand{\hlcolor}{\color}
\ifnum\version<\verpublish\else
\renewcommand{\hlcolor}[2][]{}
\fi

\ifnum\version<\verpublish
\renewcommand{\remark}[2][]{{%
  \def\emph{\textsl}\def\textbullet{$\bullet$}%
  \def\tmparga{#1}\def\tmpargb{draft}\ifx\tmparga\tmpargb%
  \color[rgb]{0.5,0,0}\normalfont\sffamily\hspace{1ex} #2\hspace{1ex}\fi}}
\fi

\ifnum\version<\verdraft
\renewcommand{\remark}[2][]{{\normalfont\sffamily\hspace{1ex}%
  \def\emph{\textsl}\def\textbullet{$\bullet$}%
  \def\tmparga{#1}%
  \def\tmpargb{NB}\ifx\tmparga\tmpargb\color[rgb]{0,0,0.8}\fi%
  \def\tmpargb{BK}\ifx\tmparga\tmpargb\color[rgb]{0,0.5,0}\fi%
  \def\tmpargb{}\ifx\tmparga\tmpargb\color{red}\fi%
  \def\tmpargb{}\ifx\tmparga\tmpargb\else \textbf{#1:} \fi%
  #2\hspace{1ex}}}
\fi

\metaset{title}{Yangian Symmetry Escapes from the Fishnet}
\metaset{author}{Niklas Beisert, Benedikt König}

\begin{document}

\ifnum\version>0
\pdfbookmark[1]{Title Page}{title}
\thispagestyle{empty}

\begingroup\raggedleft\footnotesize\ttfamily
\arxivlink{2511.19788}%
\par\endgroup

\vspace*{2cm}
\begin{center}%
\begingroup\Large\bfseries\metapick[print]{title}\par\endgroup
\vspace{1cm}

\begingroup\scshape
\metapick[print]{author}
\endgroup
\vspace{5mm}

\textit{Institut für Theoretische Physik,\\
Eidgenössische Technische Hochschule Zürich,\\
Wolfgang-Pauli-Strasse 27, 8093 Zürich, Switzerland}
\par\vspace{1mm}
\texttt{nbeisert@itp.phys.ethz.ch}, \texttt{benedikt.koenig@aei.mpg.de}

\vfill

\metaif[]{draft}{
\vspace{5mm}
\begingroup\bfseries\Large \metaterm{draft}\endgroup
\par
\vspace{5mm}
\vfill
}{}

\textbf{Abstract}\vspace{5mm}

\begin{minipage}{12.7cm}
We investigate Yangian symmetry for the equations of motion and the action
of the classical bi-scalar and supersymmetric fishnet models in four spacetime dimensions,
and we subsequently discuss its applicability to planar correlation functions.
We argue that Yangian symmetry is classically realised in these models 
subject to specific evaluation parameter patterns.
Curiously, Yangian invariance does not extend to generic quantum correlation functions 
in the bi-scalar model beyond the well-established classes of Yangian invariant correlators.
We present several concrete counter-examples
of bi-scalar correlators given by sums of Feynman graphs
and of bi-scalar graphs with octagon-shaped loops.
This finding underlines the notion that a non-zero dual Coxeter number
represents an obstacle towards quantum Yangian symmetry
and possibly also for complete integrability in planar QFT models.
\end{minipage}

\vspace*{3cm}

\end{center}

\ifnum\version>1
\newpage
\tableofcontents
\fi

\newpage
\fi

\section{Introduction}
\label{sec:intro}

The detailed study of spectra within the AdS/CFT correspondence
led to the discovery of integrable structures for
superstrings on the $\mathrm{AdS}^5\times\mathrm{S}^5$ background and in $\superN=4$ supersymmetric Yang--Mills (SYM) theory
as well as for analogous pairs of string backgrounds and CFT models,
all in their planar limits, see the reviews \cite{Beisert:2010jr,Bombardelli:2016rwb}.
This planar integrability extended earlier observations of integrable structures within 
certain sectors of Yang--Mills and QCD models
in the 't~Hooft limit of a large number of colours 
\cite{Faddeev:1994zg}.
Integrability appears to be an exact feature
for planar $\superN=4$ SYM in four spacetime dimensions,
valid to all perturbative orders and applicable to wide classes of observables.
Lacking a clear-cut definition, a proof of integrability remains elusive;
nevertheless, the integrability feature has, thus far, withstood all tests.

In general, integrability is a delicate feature which relies on fine-tuning of its parameters. 
However, once established, it typically proves surprisingly robust
enduring quantum effects at all perturbative orders including the procedure of renormalisation.
This resilience certainly applies to $\superN=4$ SYM
whose fields and interaction terms are all well-balanced to ensure 
a large amount of supersymmetry paired with conformal symmetry.
Its planar integrability functions at all orders as well as non-perturbatively,
and it has enabled the computation of a broad set of data using integrability-based techniques,
both at high perturbative orders around weak coupling and non-perturbatively at large and finite coupling.

In view of this delicacy, it came as a surprise that 
there exists a drastically simplified variant of a planar QFT model
displaying signs of the same kind of integrability:
This so-called bi-scalar model describes the dynamics
of two complex matrix-valued scalar fields
interacting via a particular quartic term \cite{Gurdogan:2015csr}.
On the formal level, this model disregards the fundamental QFT axiom of unitarity;
nonetheless, this evident shortcoming does not preclude investigations and calculations.
On the positive side, the model is classically conformal,
and conformal symmetry is maintained by planar quantum effects
because there are (almost) no planar divergences.
These features are not entirely unexpected, 
as the model can be obtained as a limit \cite{Gurdogan:2015csr,Caetano:2016ydc}
of an integrability-preserving twist deformation of $\superN=4$ SYM
\cite{Frolov:2005ty,Frolov:2005dj,Frolov:2005iq}.
And, in fact, substantial evidence towards integrability of the bi-scalar model has been found
\cite{Gurdogan:2015csr,Caetano:2016ydc,Chicherin:2017cns,Gromov:2017cja,Chicherin:2017frs}.

Much of the simplicity and many surprising features of the bi-scalar model
are founded on the particular flavour-chiral nature of its quartic interaction term:
The interactions allow only a specific ordering of flavours
which in the planar limit forces the flavour lines
to follow a very regular pattern of parallel lines intersecting at right angles;
its planar Feynman graphs have a typical “fishnet” (square lattice) structure
which lends the class of models its name.
Interestingly, fishing-net Feynman graphs appeared much earlier
as an integrable and thus solvable system \cite{Zamolodchikov:1980mb}.
The planar anomalous dimensions of local operators in this CFT
receive contributions from graphs with fishnet-like structures.
Not only can anomalous dimensions be obtained from integrability-based spectral curve techniques,
but even the contributing Feynman diagrams can be computed iteratively to some extent
\cite{Gurdogan:2015csr,Caetano:2016ydc,Gromov:2017cja,Ipsen:2018fmu}, see \cite{Kazakov:2018ugh} for a review.

Integrability for planar correlation functions
(of individual fields at distinct spacetime points and
whose colour degrees of freedom are arranged in a single overall trace)
is expressed in terms of invariance relations, so-called Ward--Takahashi identities,
under an extended symmetry structure known as a Yangian algebra \cite{Drinfel'd:1985,Drinfel'd:1988}.
In this context, the Yangian algebra extends the conformal symmetries
which are realised as differential relations for the correlation functions
and which imply that their spacetime dependency must be solely through conformal cross ratios.
The Ward--Takahashi identities for Yangian symmetry
have a similar differential form as their conformal counterparts,
and they impose additional constraints on the dependency 
of the planar correlation function on the conformal cross ratios.

Large classes of planar graphs for correlation functions 
were shown to be invariant under the conformal Yangian algebra
\cite{Chicherin:2017cns,Chicherin:2017frs,Kazakov:2023nyu,Loebbert:2025abz}.
An essential part of this deduction consisted in establishing an appropriate
representation for the Yangian generators.
The Yangian representation is determined largely by its coalgebra,
but it allows so-called evaluation parameters
as continuous degrees of freedom for all external legs.
It turns out that these evaluation parameters must be adjusted
to the structure of the graph to render it properly invariant.
Furthermore, loops must follow a specific structure in order
to maintain Yangian invariance.
Two widely accepted assumptions concerning planar correlation functions in the bi-scalar model,
namely that they are represented by a single unique Feynman graph, 
see e.g.\ \cite{Chicherin:2017cns,Loebbert:2020tje,Chicherin:2022nqq}
and that all loops have a square shape,
ensure that Yangian invariance applies to the correlator:
Uniqueness ensures that there is only a single graph structure
to determine the applicable evaluation parameters,
and square loops satisfy a structural constraint derived from Yangian symmetry.
Using Yangian invariance allowed to construct 
a functional basis for selected correlation functions in the bi-scalar model
and the applicable linear combination was selected by
achieving an appropriate singularity structure for the Feynman graph
\cite{Loebbert:2019vcj,Corcoran:2020epz}.

\medskip

In the present work, we will address two questions
regarding Yangian symmetry for the aforementioned bi-scalar model
as well as a supersymmetric extension of it:
Does the Yangian invariance apply in an elementary way
as a symmetry for the equations of motion and the action of these models?
Does this classical symmetry extend further to correlation functions
of these models?

Similar questions have been asked for their $\superN=4$ SYM ancestor model
and evaluated positively \cite{Beisert:2017pnr,Beisert:2018zxs,Beisert:2018ijg}.
Furthermore, the twist deformation preserves essential parts of the Yangian symmetry \cite{Garus:2017bgl,Beisert:2024wqq}.
However, the limit leading to the fishnet models is rather singular,
and could conceivably impair the symmetry.
It turns out that Yangian invariance in the fishnet models
requires the use of non-trivial evaluation parameters,
which is an additional feature compared to the Yangian invariance in the ancestor models.
A central goal of the present work is to understand
the appropriate implementation of these evaluation parameters
for correlation functions in fishnet models.

The appearance of evaluation parameters can be linked to the reduction of Yangian symmetry
from the underlying superconformal algebra $\alg{psu}(2,2|4)$ for $\superN=4$ SYM
to the (super)conformal algebras $\alg{su}(2,2|\superN)$
for the reduced models with $\superN=0,1$ supersymmetries.
A key purpose of the evaluation parameters is to make the Yangian representation 
compatible with the cyclic structure of planar correlation functions
which have a disk-like planar topology.
Yangian representations typically do not respect this cyclic structure
thus spoiling the possibility of consistent invariance statements for planar correlators
\cite{Drummond:2009fd}.
Interestingly, these cyclicity-violating effects
can be compensated by a suitable arrangement of evaluation parameters
with a quasi-periodic shift for each cycle around the planar disk
\cite{Chicherin:2017cns,Chicherin:2017frs}. 
The magnitude of this shift must match the dual Coxeter number $h^*=4-\superN$
of the underlying (super)conformal algebra $\alg{su}(2,2|\superN)$
in order to repair cyclicity.
In the fishnet models with $\superN=0,1$ this quasi-periodicity
clearly requires a non-trivial assignment of evaluation parameters,
which is unnecessary for the $\superN=4$ SYM ancestor model
with its vanishing dual Coxeter number $h^*=0$.

\medskip

The present work is organised as follows:
We start in \secref{sec:models} by introducing the two fishnet models of interest
and their Yangian algebra structures.
In \secref{sec:classical}, we show that their equations of motion and actions
are invariant under Yangian symmetry
subject to appropriately chosen evaluation parameters.
In the following \secref{sec:correlators},
we address Yangian invariance of correlation functions
in the bi-scalar model
and find that it is spoiled for certain classes 
by the non-uniqueness of contributing graphs
and by incompatible loop structures.
In \secref{sec:limit}, we discuss the reduction of
Yangian symmetry from the $\superN=4$ SYM ancestor model
to the fishnet models.
We conclude in \secref{sec:conclusions} that
the evaluation parameters cannot be extended consistently to all correlation functions,
and we argue further that a non-vanishing dual Coxeter number
represents an actual obstacle towards complete integrability in these models.
Finally, in \appref{sec:uniqueness},
we discuss and show how far the uniqueness of graphs in the bi-scalar model actually applies.

\section{Fishnet Models and Yangian Representations}
\label{sec:models}

In this work, we will consider two fishnet models in four spacetime dimensions,
the non-supersymmetric bi-scalar model \cite{Gurdogan:2015csr}
and the $\superN=1$ supersymmetric brick-wall fishnet model \cite{Caetano:2016ydc}.
\unskip\footnote{Similar brick-wall models with different amounts of 
supersymmetry and/or in different spacetime dimension
have been introduced in \cite{Caetano:2016ydc,Pittelli:2019ceq,Kazakov:2018qbr,Kade:2024lkc};
we shall consider exclusively the $D=4$, $\superN=1$ supersymmetric brick-wall model in this work.}
For simplicity, we shall refer to these models as the “bi-scalar model” 
and the “brick-wall model”,
respectively.
The bi-scalar model has merely two complex scalar fields $\phi^1, \phi^2$
taking values in matrices of some dimension $N_{\text{c}}$.
\unskip\footnote{We shall be interested in the planar limit $N_{\text{c}}\to\infty$,
and thus the fields will always be treated as abstract matrices
whose ordering within field polynomials does matter.
Furthermore, we will disregard (higher-loop) renormalisation effects
related to wrapping \cite{Sieg:2016vap,Grabner:2017pgm}.}
The action reads:
\unskip\footnote{We fix the coupling constant to some value
noting that it may be recovered by a simultaneous rescaling
of the fields and the action.}
\[@{eq:biscalar}
\action_{\superN=0}
=
\int (\diff x)^4 \tr \brk[s]!{
-\partial^\mu\bar\phi^1\.\partial_\mu\phi^1
-\partial^\mu\bar\phi^2\.\partial_\mu\phi^2
-\bar\phi^1\bar\phi^2\phi^1\phi^2
}
\]
Notably, the model has the quartic scalar interaction term $\bar\phi^1\bar\phi^2\phi^1\phi^2$
but excludes its conjugate $\bar\phi^2\bar\phi^1\phi^2\phi^1$
thus rendering the action complex and the model non-unitary.

The supersymmetric brick-wall model consists of three
chiral supermultiplets $\Phi^a=(\phi^a,\psi^a)$, $a=1,2,3$,
which interact via the chiral and anti-chiral pre-potentials 
$\tr(\Phi^1\Phi^2\Phi^3)$ and $\tr(\bar\Phi^1\bar\Phi^2\bar\Phi^3)$,
but not their respective conjugates
$\tr(\bar\Phi^3\bar\Phi^2\bar\Phi^1)$ and $\tr(\Phi^3\Phi^2\Phi^1)$.
The action in components reads
\unskip\footnote{The flavour index $a$ is implicitly summed over $a=1,2,3$
and considered modulo 3.}
\<@{eq:brickwall}
\action_{\superN=1}
\&?\simeq
\int (\diff x)^4 \tr \brk[s]^2[
-\partial^\mu\bar\phi^a\.\partial_\mu\phi^a
+
\bar\psi^a\gamma{\cdot}\partial\psi^a
\vspace{-0.5ex}\\\vspace{-0.5ex}\&;\qquad\qquad\quad
+
\phi^a\psi^{a+1}{\cdot}\psi^{a-1}
+
\bar\phi^a\bar\psi^{a+1}{\cdot}\bar\psi^{a-1}
-
\bar\phi^{a}\bar\phi^{a+1}\phi^{a}\phi^{a+1}
\brk]
\>
Supersymmetry implies a unique form for all terms involving fermions.
For reasons of conciseness,
we shall sketch the form of these terms throughout this work 
rather than exposing details such as their spinor indices.
Concrete examples and calculations will be given
in terms of the bi-scalar model.
Brick-wall extensions of the resulting expressions
will be sketched by highlighting the applicable generalisations.

\medskip

The above models enjoy conformal and $\superN=1$ superconformal symmetry, respectively, 
which are described by the algebras $\alg{su}(2,2)$ and $\alg{su}(2,2|1)$.
The conformal algebra is spanned by the spacetime rotations $\gen{L}_{\mu\nu}$, 
the translations $\gen{P}_\mu$, the scaling $\gen{D}$ 
and the conformal boosts $\gen{K}_\mu$.
We normalise the conformal generators
and their representation on the scalar fields as:
\<
\gen{P}^\mu \phi \&= \iunit \partial^\mu \phi
\&
\gen{K}^\mu \phi \&= 
\iunit x^\mu x^\nu \partial_\nu \phi
-\ihalf x^2 \partial^\mu \phi
+\iunit x^\mu \phi
\\
\gen{L}^{\mu\nu} \phi \&= 
-\iunit x^\mu \partial^\nu \phi
+\iunit x^\nu \partial^\mu \phi
\&
\gen{D} \phi \&= 
\iunit x^\nu \partial_\nu \phi
+\iunit \phi
\>
The superconformal generalisation additionally has
supersymmetry translations $\gen{Q}$, $\gen{\bar Q}$,
superconformal boosts $\gen{S}$, $\gen{\bar S}$
and the internal rotation $\gen{R}$.
Their representation on the fields is uniquely determined,
but we will not expose its explicit form.

\medskip

The (super)conformal algebra can be enhanced to a Yangian algebra
by adding one “level-one” generator $\genyang{J}^A$
for each (super)conformal “level-zero” generator $\gen{J}^A$.
Supposing the conformal algebra takes the general form
$\comm{\gen{J}^A}{\gen{J}^B}=\iunit F^{AB}{}_C\gen{J}^C$
and the quadratic invariant takes the form $C_{AB}\gen{J}^A\gen{J}^B$,
the level-one generators act as the bi-local combination:
\[@{eq:levelone},
\genyang{J}^C \simeq \ihalf F^C{}_{AB} (\gen{J}^A \wedge \gen{J}^B)
\],
where $F^C{}_{AB}\defeq F^{CD}{}_B C_{DA}$ denotes a dual set of structure constants.
We will describe the action of bi-local terms $\gen{J}^A \wedge \gen{J}^B$ further below.
Concretely, the level-one momentum in the non-supersymmetric case
takes the form:
\[
\genyang{P}^\mu_{\superN=0}
\simeq
\iunit\.\gen{L}^\mu{}_\nu\wedge\gen{P}^\nu
+\iunit\.\gen{D}\wedge\gen{P}^\mu
\]
For the $\superN=1$ supersymmetric case, there is an additional term
of the form $\gen{Q}\wedge\gen{\bar Q}$
whose form is uniquely determined but will not be exposed either.

\section{Classical Yangian Invariance}
\label{sec:classical}

We can now discuss Yangian invariance of the classical fishnet models
along the lines of \cite{Beisert:2017pnr,Beisert:2018zxs}.
\unskip\footnote{We shall not provide a detailed introduction to
the framework for classical Yangian symmetry for planar gauge theory models,
which is explained at length in \cite[section 3 and 4]{Beisert:2018zxs}
and in \cite[section 2]{Beisert:2018ijg}.}
These models have conformal symmetry, hence it remains to show
invariance under the level-one Yangian symmetries.

Let us start with the equations of motion for the bi-scalar model.
They all take the common form
\[@{eq:biscalareom}
\frac{\var S}{\var \phi} = \partial^2\phi - \phi\phi\phi
\],
where the conjugation and flavour configuration of the scalar fields
does not actually play a role towards the conformal representations.
The level-one generators \eqref{eq:levelone}
act on the field polynomials $\oper{X}_{\nfield{L-1}}$
constituting the equations of motions as:
\[@{eq:levelonerep}
\genyang{J}^C \oper{X}_{\nfield{L-1}} 
=
\iunit F^C{}_{AB}
 \sum_{j<k=1}^{L-1} \gen{J}^A_j \gen{J}^B_k \oper{X}_{\nfield{L-1}}
+ \sum_{k=1}^{L-1} \genyang{J}^C_{L,k} \oper{X}_{\nfield{L-1}}
\]
Here, the operator $\gen{J}^C_{k}$ describes the action of the level-zero generator $\gen{J}^C$
on the $k$-th field of the polynomial $\oper{X}_{\nfield{L-1}}$,
whereas $\genyang{J}^C_{L,k}$ describes a single-field action of the level-one generator $\genyang{J}^C$ on the $k$-th field.
The latter is a natural component of a Yangian representation
where it may be regarded as a short-distance completion of the bi-local terms. 
We emphasise that we take $L$ to refer to the number of fields in the term $\action_{\nfield{L}}$ of the action
which gives rise to the corresponding term $\oper{X}_{\nfield{L-1}}=\var \action_{\nfield{L}}/\var \phi$ 
in the equation of motion of length $L-1$.
This notational shift by $1$ will make our later comparison to invariance of the action more immediate and convenient.

Let us construct the action of the level-one momentum $\genyang{P}^\mu_{\superN=0}$.
Its bi-local contribution on any ordered pair of scalar fields $\phi\otimes\phi$ is given as:
\<@{eq:Pphiphi}
\&:
\genyang{P}^\mu_{\superN=0,\bilocal}(\phi\otimes\phi)
\\=
\iunit(\gen{L}^\mu{}_{\nu} \phi)\otimes(\gen{P}^\nu \phi)
+\iunit(\gen{D} \phi)\otimes(\gen{P}^\mu \phi)
-\iunit(\gen{P}^\nu \phi)\otimes(\gen{L}^\mu{}_{\nu} \phi)
-\iunit(\gen{P}^\mu \phi)\otimes(\gen{D} \phi)
\\=
(\iunit\partial^\mu\phi)\otimes\phi-\phi\otimes(\iunit\partial^\mu\phi)
=(\gen{P}^\mu_1-\gen{P}^\mu_2)(\phi\otimes\phi)
\>
It is now straight-forward to compose the level-one action \eqref{eq:levelonerep}
on the sequence of three scalar fields $\phi\phi\phi$ in the equations of motions \eqref{eq:biscalareom}
as the sum of bi-local terms on all pairs of fields \eqref{eq:Pphiphi}
plus a yet-unspecified local contribution $\genyang{P}^\mu_{4,j}$ on all fields:
\[
\genyang{P}^\mu_{\superN=0}(\phi\phi\phi)
=
\brk!{2\gen{P}^\mu_1-2\gen{P}^\mu_3+\genyang{P}^\mu_{4,1}+\genyang{P}^\mu_{4,2}+\genyang{P}^\mu_{4,3}}(\phi\phi\phi)
\]
Conversely, the kinetic term $\partial^2\phi$ in the equations of motion contributes
only via the single-field action $\genyang{P}^\mu_{2,1}$.
From this form we can infer a suitable single-field action of $\genyang{P}^\mu_{L,k}$
which renders the equations of motion invariant. 
The single-field action takes the form
($L=2,4$ refers to the length of the corresponding action term):
\[@{eq:singlefieldbi-scalar}
\genyang{P}^\mu_{\superN=0;L,k} = -4\frac{L-2k}{L}\gen{P}^\mu_k
\]
Some comments on this form are in order: 
Equating the single-field action of a level-one generator
with the corresponding level-zero generator is known as an evaluation representation.
The prefactor serves as the evaluation parameter for the representation,
and a dependency on the tensor site $k$ is perfectly acceptable.
Here, the evaluation parameters further depend on the overall length $L-1$
of the field monomial term, which may seem unusual,
but so is a Yangian representation on polynomials with mixed lengths.
In fact, the dependency of the evaluation parameters on the site is linear and anti-symmetric.
Furthermore, the dependency is in line with observations for correlators,
see \secref{sec:correlators}.
Finally, we have checked explicitly
that covariance of the equations of motion holds for all level-one Yangian generators
with the same prescription for the evaluation parameters:
\[
\genyang{J}^C_{\superN=0} \frac{\var S}{\var\phi}=0
\]

\smallskip

Let us now consider invariance of the action,
which ought to follow by integration of the covariance statement for the equations of motion.
To that end, we have verified that the covariance statement is a closed variational form,
see \cite{Beisert:2024wqq} for details,
provided that the single-field action takes the form \eqref{eq:singlefieldbi-scalar};
consequently, the covariance statement has a unique indefinite integral
serving as the invariance statement for the action.

Invariance of the action turns out to be almost automatic in this simple case:
All the bi-local contributions reduce to differences as in \eqref{eq:Pphiphi},
or to terms proportional to conformal transformations.
Importantly, their action does not depend on the scalar field flavours 
which is the only aspect that distinguishes the fields within the trace.
\unskip\footnote{We acknowledge earlier discussions with F.\ Loebbert on this issue.}
\footnote{For example, the level-one momentum acts on any pair of scalar fields
as in \eqref{eq:Pphiphi}. The flavours of the scalar fields are insignificant
because the bi-local action neither depends on them
nor does it change their ordering.
Upon using invariance under cyclic permutations
(modulo flavours, which play no role),
all single-momentum terms $\gen{P}_k$ become equivalent.
Differences of individual momenta therefore cancel,
and any residual momentum contribution from the evaluation parameters
averages to the total momentum $\gen{P}$
which is zero due to translation invariance.}
Upon using cyclic symmetry of the action, 
all resulting terms therefore cancel.
We have verified the cancellation explicitly for all generators $\genyang{J}^C$.
Altogether, the action is invariant under the level-one generators:
\[
\genyang{J}^C_{\superN=0} \action_{\superN=0}=0
\]

\medskip

Next, let us consider the supersymmetric case
which will require far more elaborate cancellations.
The equations of motion turn out to be covariant
provided that the evaluation parameters are adjusted as:
\[@{eq:singlefieldsusy}
\genyang{J}^C_{\superN=1;L,k} = -\brk[s]*{3\frac{L-2k}{L}\pm \half\delta_{L,4}(\delta_{k,1}+\delta_{k,3}) }\gen{J}^C_k
\]
Several comments are in order:
First, the dependency of the evaluation parameter 
on the length and on the site is analogous to the bi-scalar model. 
Merely, the overall coefficient has changed from $4$ to $3$.
This adjustment is related to the dual Coxeter number $h^*$ of the underlying conformal algebra
which appears as the algebraic product of the two sites
of the bi-local terms in the level-one representation \eqref{eq:levelone}:
\[@{eq:dualcoxdef}
[\genyang{J}^C] \defeq
\ihalf F^C{}_{AB} \comm{\gen{J}^A}{\gen{J}^B}
=
-\half F^C{}_{AB} F^{AB}{}_D\gen{J}^D
=-h^* \gen{J}^C
\]
For the conformal algebra $\alg{su}(2,2)$, it equals $h^*=4$, e.g.
\[
[\genyang{P}^\mu_{\superN=0}] =
\iunit\comm{\gen{L}^\mu{}_\nu}{\gen{P}^\nu}
+\iunit\comm{\gen{D}}{\gen{P}^\mu}
=-4\gen{P}^\mu
\],
whereas for the superconformal algebra $\alg{su}(2,2|1)$, this number reduces to $h^*=3$
due to the supersymmetry contribution of the form $\acomm{\gen{Q}}{\gen{\bar Q}}$.
In essence, the combinations $4\gen{J}^C$ and $3\gen{J}^C$
both incorporate $-[\genyang{J}^C]$ for the respective models.
Second, the covariance relation for the equations of motion
$\genyang{J}^C (\var S/\var\Phi)=*(\var S/\var\Phi)\simeq 0$
holds modulo all equations of motion.
The coefficients `$*$' for the equation of motion terms $\var S/\var\Phi$
on the right-hand side
follow the form as described in \cite{Beisert:2018zxs}.
\unskip\footnote{There is a minor deviation 
when acting with the fermionic level-one generators
on the fermionic equations of motion.
The extra terms can be traced to the algebra relations 
in the combination $[\genyang{J}^C]$ in \eqref{eq:dualcoxdef}
which do not hold strictly,
but only modulo the fermionic equations of motion.}
Third, on top of the linear dependency of the evaluation parameters \eqref{eq:singlefieldsusy} on the site $k$,
there is also an extra pattern when acting on 
the cubic products of scalars in the scalar equations of motion. 
For $\var S/\var\bar\phi$, these are of the kind 
$\phi\phi\bar\phi$ and $\bar\phi\phi\phi$,
whereas for $\var S/\var\phi$
they take the form $\bar\phi\bar\phi\phi$ or $\phi\bar\phi\bar\phi$.
The sign of the evaluation parameter shift curiously depends on 
whether the first two fields are of equal (plus) or of opposite (minus) conjugation type.
We will discuss this term in more depth later.

Let us finally turn to invariance of the action.
To that end, we use the cyclicity-respecting level-one representation $\genyang{J}'$
proposed in \cite{Beisert:2018zxs} and elaborated in \cite{Beisert:2024wqq}.
Here, the strength of the bi-local contributions depends on the distance of the two sites 
such that the relative weight interpolates linearly between $+1$ and $-1$ around the trace:
\unskip\footnote{We disregard the peculiarities of the implementation 
due to length-changing terms in the supersymmetry representation.
See \cite{Beisert:2018zxs} for details on the implementation.}
\[
\genyang{J}'^C \oper{X}_{\nfield{L}} 
=
 \sum_{k=1}^L \genyang{J}'^C_k \oper{X}_{\nfield{L}}
+
\ihalf F^C{}_{AB}
 \sum_{j=1}^L\sum_{k=1}^{L-1} \frac{L-2k}{L} \gen{J}^A_j \gen{J}^B_{j+k} \oper{X}_{\nfield{L}}
\]
This form is equivalent to the non-manifestly cyclic variant $\genyang{J}^C$
via the relation:
\[
\genyang{J}'^C \oper{X}_{\nfield{L}} 
=
\genyang{J}^C \oper{X}_{\nfield{L}}
+\sum_{k=1}^L\brk[s]*{\genyang{J}'^C_k-\genyang{J}^C_k+\frac{L-2k}{L} [\genyang{J}^C]_k} \oper{X}_{\nfield{L}}
-\iunit F^C{}_{AB} \sum_{k=1}^{L} \frac{L-2k}{L} \gen{J}^A_k (\gen{J}^B \oper{X}_{\nfield{L}})
\]
The last term is irrelevant when acting on the superconformally invariant action
because $\gen{J}^B S=0$. 
The middle term describes the appropriate transformation for the single-site terms.
We note that the shift contribution equals precisely the form of \eqref{eq:singlefieldbi-scalar}
and the first term in \eqref{eq:singlefieldsusy}.
It therefore suffices to set
\unskip\footnote{We have added a contribution independent of the site
which is equivalent to the overall level-zero representation.
This term is essential for the length-changing terms
in the supersymmetry representation.}
\[
\genyang{J}'^C_{\superN=0;L,k} = 0
\eqnjoin{\text{and}}
\genyang{J}'^C_{\superN=1;L,k} = \pm \rfrac{1}{4}\delta_{L,4}(-1)^{k-1} \gen{J}^C_k
\]
The sign in the $\superN=1$ case is minus when the first two scalar fields are of equal conjugation type,
otherwise it is plus; more explicitly:
\[@{eq:anomsusy}
\genyang{J}'^C_{\superN=1,\local} (\bar\phi^{a}\bar\phi^{a+1}\phi^{a}\phi^{a+1})
=
\rfrac{1}{4}\brk!{-\gen{J}^C_1+\gen{J}^C_2-\gen{J}^C_3+\gen{J}^C_4}
(\bar\phi^{a}\bar\phi^{a+1}\phi^{a}\phi^{a+1})
\]
The remaining local contribution thus incorporates a particular choice
of evaluation parameters for the quartic term in the action.
Notably, this configuration of evaluation parameters
depends on the conjugation type pattern.
Altogether, we have verified concretely that all level-one Yangian generators leave the action invariant 
using the cyclic prescription in \cite{Beisert:2018zxs}
and taking into account the extra contributions due to evaluation parameters in \eqref{eq:anomsusy}.
Whether or not \eqref{eq:anomsusy} describes an admissible and/or desirable (almost) local term 
for the action of a Yangian generator may be debated.
Here we merely state that this term is required to achieve invariance for the classical action.

\section{Yangian Representation on Correlators}
\label{sec:correlators}

A symmetry of the action typically implies invariances
for correlation functions; these are known as Ward--Takahashi identities.
Let us therefore study Yangian invariance
for planar correlation functions of the bi-scalar model 
which we shall denote as “bi-scalar correlators” for short.
These correlators are composed from planar Feynman graphs
which have a connected planar disk-like topology 
and which we shall call “bi-scalar graphs”.
A representation of the conformal Yangian on the class of so-called “fishnet graphs”
has been introduced in \cite{Chicherin:2017cns,Chicherin:2017frs}.
Fishnet graphs form the subset of bi-scalar graphs
which can be cut out from a complete square fishnet grid by cutting some links.
With a suitable arrangement of evaluation parameters for the external legs,
it has been shown that fishnet graphs are invariant under Yangian generators.
This Yangian invariance is actually a part of a broader class of loom graphs \cite{Kazakov:2023nyu}
and rather general planar scalar Feynman graphs \cite{Loebbert:2025abz}
(see also references therein).
A core feature of this representation is that the arrangement of evaluation parameters
depends on the graph topology.
In order to establish Yangian invariance for bi-scalar correlators,
we need to describe a suitable arrangement of evaluation parameters
based on the given sequence of external legs for the correlator.
Ideally, such a transformation rule would be deduced by somehow composing 
the evaluation parameters used for the equations of motion and the action;
however, this approach would not work out as we shall see.

\medskip

Let us first recollect the pertinent results for Yangian invariance of fishnet graphs,
see \cite{Chicherin:2017cns,Chicherin:2017frs} for further details:
For convenience, we label the four flavours of scalar fields $(\phi_1,\phi_2,\bar\phi_1,\bar\phi_2)$
as the $\Integer_4$-integers $(1,2,3,4)\in\Integer_4$.
A fishnet graph is thus labelled by its $\Integer_4$-integer sequence $f_j$ of external legs
along the perimeter of the graph.
Furthermore, we denote the number of propagators connecting two adjacent legs $j$ and $j+1$ 
along the perimeter of the graph by $d_j$ discounting for the external legs themselves.
The evaluation parameters $s_j$ for the legs
are then specified by the difference relation 
\[
s_{j+1}-s_j=1-d_j
\]
An overall shift of all evaluation parameters is irrelevant,
and it makes sense to fix $s_1$ to match $f_1$ modulo 4;
then the evaluation parameters $s_j$ modulo 4 turn out to 
reproduce the sequence of particle flavours $f_j$
due to the particular flavour arrangement of the vertices.
It has been shown in \cite{Chicherin:2017cns,Chicherin:2017frs}
that the fishnet graph is invariant under the conformal Yangian representation \eqref{eq:levelonerep}
with the single-site action $\genyang{J}_j=2 s_j\gen{J}_j$
and the given arrangement of evaluation parameters $s_j$.

\begin{figure}\centering
\begin{mpostfig}
interim xu:=1.5cm;
paths[1]:=(0xu,-0.25xu)--(0xu,1.25xu);
paths[2]:=(1xu,-0.25xu)--(1xu,1.25xu);
paths[3]:=(2xu,-0.25xu)--(2xu,0.25xu);
paths[4]:=(2xu, 0.75xu)--(2xu,1.25xu);
paths[5]:=(-0.25xu,1xu)--(2.25xu,1xu);
paths[6]:=(-0.25xu,0xu)--(2.25xu,0xu);
paths[8]:=(2.2xu,1.325xu){dir -135}..(2xu,1.125xu){dir 180}--(1xu,1.125xu)--(0xu,1.125xu)
        {dir 180}..{dir 270}(-0.125xu,1xu)--(-0.125xu,0xu)
        {dir -90}..{dir 0}(0xu,-0.125xu)--(2xu,-0.125xu)
        {dir 0}..{dir 90}(2.125xu,0xu)..{dir 180}(2xu,0.125xu)--(1.25xu,0.125xu)
        {dir 180}..{dir 90}(1.125xu,0.25xu)--(1.125xu,0.75xu)
        {dir 90}..{dir 0}(1.25xu,0.875xu)--(2.0xu,0.875xu)
        {dir 0}..{dir 90}(2.125xu,1xu)
        {dir 90}..{dir 45}(2.325xu,1.2xu);
drawtwo paths[1];
drawtwo paths[2];
drawtwo paths[3];
drawtwo paths[4];
drawone paths[5];
drawone paths[6];
draw (2.15xu,1.15xu)--(2.4xu,1.4xu) pensize(1pt) dashed evenly withcolor red;
draw paths[8] dashed evenly withcolor 0.5[red,black];
arrowtwo (reverse paths[1], 0.5);
arrowtwo (reverse paths[2], 0.5);
arrowtwo (reverse paths[3], 0.75);
arrowtwo (reverse paths[4], 0.75);
arrowone (reverse paths[5], 0.3);
arrowone (reverse paths[6], 0.3);
label.top(btex 1 etex, point 1 of paths[4]);
label.top(btex 1 etex, point 1 of paths[2]);
label.top(btex 1 etex, point 1 of paths[1]);
label.lft(btex 2 etex, point 0 of paths[5]);
label.lft(btex 2 etex, point 0 of paths[6]);
label.bot(btex 3 etex, point 0 of paths[1]);
label.bot(btex 3 etex, point 0 of paths[2]);
label.bot(btex 3 etex, point 0 of paths[3]);
label.rt (btex 4 etex, point 1 of paths[6]);
label.top(btex 5 etex, point 1 of paths[3]);
label.bot(btex 3 etex, point 0 of paths[4]);
label.rt (btex 4 etex, point 1 of paths[5]);
\end{mpostfig}
\caption{A unique fishnet graph depicted with a suitable sequence 
of evaluation parameters along the perimeter curve
which has been cut open between the two legs at the upper right corner.}
\label{fig:unique}
\end{figure}

The remaining tasks towards establishing Yangian invariance of bi-scalar correlators
are to find the sequence of evaluation parameters $(s_i)$
corresponding to a flavour configuration $(f_i)$ of the external legs,
and to show that all contributing bi-scalar graphs
are invariant under this Yangian representation.
Let us mention some relevant properties based on the fishnet graph topology:
First, there can be no more than three consecutive propagators between two external legs, $d_j\leq 3$. 
Therefore, $s_{j+1}$ is constrained to lie between $s_j-2$ and $s_j+1$.
Further, the evaluation parameter matches the flavour number, $s_{j+1}\equiv f_{j+1}\pmod 4$, 
which determines $s_{j+1}$ from $s_j$ unambiguously.
Second, the sequence of $s_j$ increases by $\Delta s=+4$ for a cycle around the graph 
which indicates one complete flavour cycle around the perimeter.
Third, all internal loops are squares.
Altogether, it is straight-forward to construct the evaluation parameter sequence $(s_j)$ 
for a given flavour sequence $(f_j)$ of a fishnet graph,
see \figref{fig:unique} for an example.
Its flavour and resulting evaluation parameter sequences are given by:
\<@{eq:unique}
f_{6,0} \&= (1,1,1,2,2,3,3,3,4,1,3,4)
\\
s_{6,0} \&= (1,1,1,2,2,3,3,3,4,5,3,4)^{+4}
\>
We display the quasi-periodicity $\Delta s=+4$ of the evaluation parameter sequence
as a superscript because it is a relevant piece of information.
The fishnet graph topology can be reconstructed by tracing out the sequence of flavours $(f_j)$
on a complete fishnet and cutting out the enclosed fishnet graph.
A hallmark feature of the bi-scalar model is that
this graph turns out to be the unique contribution to the planar correlator specified
by this sequence of flavours.
We provide arguments towards uniqueness of fishnet and more general bi-scalar graphs in \appref{sec:uniqueness}.

\medskip

Importantly, not all contributions to bi-scalar correlators are of fishnet graph type:
Such graphs have been described in \cite{Chicherin:2017cns} as “singular fishnet graphs”. 
They are cut out from a fishnet plane with a non-trivial structure of interconnected layers or sheets
in analogy to the Riemann sheets of complex analysis.
In other words, if the graph is brought to a square lattice arrangement, 
some stitches and threads will reside on top of each other.
Unlike for fishnet graphs, the corresponding perimeter curve self-intersects.
As a result, there can be more than three propagators connecting two adjacent legs
along the perimeter.
\unskip\footnote{On a regular fishnet graph, this would necessarily close a loop,
but here the layered fishnet allows stacking of stitches.}
This class of graphs has two adverse features towards Yangian symmetry:
First, a non-fishnet graph is not necessarily the unique contribution 
to the corresponding bi-scalar correlator.
\unskip\footnote{There have been various claims that all bi-scalar correlators
are given by unique Feynman graphs,
see e.g.\ \cite{Chicherin:2017cns,Loebbert:2020tje,Chicherin:2022nqq}. 
This confusion may have been caused by a very similar terminology
for two closely related yet different concepts.}
Second, non-fishnet graphs may involve loops of more than 4 sides.
Let us provide examples for both issues.

\medskip

\begin{figure}\centering
\begin{mpostfig}
interim xu:=1.5cm;
paths[1]:=(0xu,-0.25xu)--(0xu,1.25xu);
paths[2]:=(1xu,-0.25xu)--(1xu,1.25xu);
paths[3]:=(2xu,-0.25xu)--(2xu,0.25xu);
paths[4]:=(3xu,-0.25xu)--(3xu,1.25xu);
paths[5]:=(-0.25xu,1xu)--(3.25xu,1xu);
paths[6]:=(-0.25xu,0xu)--(1.25xu,0xu);
paths[7]:=(1.75xu,0xu)--(3.25xu,0xu);
drawtwo paths[1];
drawtwo paths[2];
drawtwo paths[3];
drawtwo paths[4];
drawone paths[5];
drawone paths[6];
drawone paths[7];
draw (3.15xu,1.15xu)--(3.4xu,1.4xu) pensize(1pt) dashed evenly withcolor red;
arrowtwo (reverse paths[1], 0.5);
arrowtwo (reverse paths[2], 0.5);
arrowtwo (reverse paths[3], 0.8);
arrowtwo (reverse paths[4], 0.5);
arrowone (reverse paths[5], 0.5);
arrowone (reverse paths[6], 0.5);
arrowone (reverse paths[7], 0.5);
label.top(btex 1 etex, point 1 of paths[4]);
label.top(btex 1 etex, point 1 of paths[2]);
label.top(btex 1 etex, point 1 of paths[1]);
label.lft(btex 2 etex, point 0 of paths[5]);
label.lft(btex 2 etex, point 0 of paths[6]);
label.bot(btex 3 etex, point 0 of paths[1]);
label.bot(btex 3 etex, point 0 of paths[2]);
label.rt (btex 4 etex, point 1 of paths[6]);
label.top(btex \textbf{1} etex, point 1 of paths[3]) withcolor 0.3[red,black];
label.lft(btex 2 etex, point 0 of paths[7]);
label.bot(btex 3 etex, point 0 of paths[3]);
label.bot(btex 3 etex, point 0 of paths[4]);
label.rt (btex 4 etex, point 1 of paths[5]);
label.rt (btex 4 etex, point 1 of paths[7]);
\end{mpostfig}
\qquad
\begin{mpostfig}
interim xu:=1.5cm;
paths[1]:=(0xu,-0.25xu)--(0xu,1.25xu);
paths[2]:=(1xu,-0.25xu)--(1xu,0.25xu);
paths[3]:=(2xu,-0.25xu)--(2xu,1.25xu);
paths[4]:=(3xu,-0.25xu)--(3xu,1.25xu);
paths[5]:=(-0.25xu,1xu)--(3.25xu,1xu);
paths[6]:=(-0.25xu,0xu)--(1.25xu,0xu);
paths[7]:=(1.75xu,0xu)--(3.25xu,0xu);
drawtwo paths[1];
drawtwo paths[2];
drawtwo paths[3];
drawtwo paths[4];
drawone paths[5];
drawone paths[6];
drawone paths[7];
draw (3.15xu,1.15xu)--(3.4xu,1.4xu) pensize(1pt) dashed evenly withcolor red;
arrowtwo (reverse paths[1], 0.5);
arrowtwo (reverse paths[2], 0.8);
arrowtwo (reverse paths[3], 0.5);
arrowtwo (reverse paths[4], 0.5);
arrowone (reverse paths[5], 0.5);
arrowone (reverse paths[6], 0.5);
arrowone (reverse paths[7], 0.5);
label.top(btex 1 etex, point 1 of paths[4]);
label.top(btex 1 etex, point 1 of paths[3]);
label.top(btex 1 etex, point 1 of paths[1]);
label.lft(btex 2 etex, point 0 of paths[5]);
label.lft(btex 2 etex, point 0 of paths[6]);
label.bot(btex 3 etex, point 0 of paths[1]);
label.bot(btex 3 etex, point 0 of paths[2]);
label.rt (btex 4 etex, point 1 of paths[6]);
label.top(btex \textbf{5} etex, point 1 of paths[2]) withcolor 0.3[red,black];
label.lft(btex 2 etex, point 0 of paths[7]);
label.bot(btex 3 etex, point 0 of paths[3]);
label.bot(btex 3 etex, point 0 of paths[4]);
label.rt (btex 4 etex, point 1 of paths[5]);
label.rt (btex 4 etex, point 1 of paths[7]);
\end{mpostfig}
\caption{A pair of non-unique non-fishnet graphs at one loop.}
\label{fig:nonuniqueloop}
\end{figure}

\begin{figure}\centering
\begin{mpostfig}
interim xu:=1.5cm;
paths[1]:=(0xu,-0.25xu)--(0xu,1.25xu);
paths[2]:=(1xu,-0.25xu)--(1xu,0.25xu);
paths[3]:=(2xu,-0.25xu)--(2xu,1.25xu);
paths[4]:=(3xu, 0.75xu)--(3xu,1.25xu);
paths[5]:=(4xu,-0.25xu)--(4xu,1.25xu);
paths[6]:=(-0.25xu,1xu)--(2.25xu,1xu);
paths[7]:=( 2.75xu,1xu)--(4.25xu,1xu);
paths[8]:=(-0.25xu,0xu)--(1.25xu,0xu);
paths[9]:=( 1.75xu,0xu)--(4.25xu,0xu);
drawtwo paths[1];
drawtwo paths[2];
drawtwo paths[3];
drawtwo paths[4];
drawtwo paths[5];
drawone paths[6];
drawone paths[7];
drawone paths[8];
drawone paths[9];
draw (4.15xu,1.15xu)--(4.4xu,1.4xu) pensize(1pt) dashed evenly withcolor red;
arrowtwo (reverse paths[1], 0.5);
arrowtwo (reverse paths[2], 0.8);
arrowtwo (reverse paths[3], 0.5);
arrowtwo (reverse paths[4], 0.8);
arrowtwo (reverse paths[5], 0.5);
arrowone (reverse paths[6], 0.5);
arrowone (reverse paths[7], 0.5);
arrowone (reverse paths[8], 0.5);
arrowone (reverse paths[9], 0.5);
label.top(btex 1 etex, point 1 of paths[1]);
label.top(btex \textbf{5} etex, point 1 of paths[2]) withcolor 0.3[red,black];
label.top(btex 1 etex, point 1 of paths[3]);
label.top(btex 1 etex, point 1 of paths[4]);
label.top(btex 1 etex, point 1 of paths[5]);
label.bot(btex 3 etex, point 0 of paths[1]);
label.bot(btex 3 etex, point 0 of paths[2]);
label.bot(btex 3 etex, point 0 of paths[3]);
label.bot(btex \textbf{3} etex, point 0 of paths[4]) withcolor 0.3[red,black];
label.bot(btex 3 etex, point 0 of paths[5]);
label.lft(btex 2 etex, point 0 of paths[6]);
label.lft(btex 2 etex, point 0 of paths[7]);
label.lft(btex 2 etex, point 0 of paths[8]);
label.lft(btex 2 etex, point 0 of paths[9]);
label.rt (btex 0 etex, point 1 of paths[6]);
label.rt (btex 4 etex, point 1 of paths[7]);
label.rt (btex 4 etex, point 1 of paths[8]);
label.rt (btex 4 etex, point 1 of paths[9]);
\end{mpostfig}
\qquad
\begin{mpostfig}
interim xu:=1.5cm;
paths[1]:=(0xu,-0.25xu)--(0xu,1.25xu);
paths[2]:=(1xu, 0.75xu)--(1xu,1.25xu);
paths[3]:=(2xu,-0.25xu)--(2xu,1.25xu);
paths[4]:=(3xu,-0.25xu)--(3xu,0.25xu);
paths[5]:=(4xu,-0.25xu)--(4xu,1.25xu);
paths[6]:=(-0.25xu,1xu)--(1.25xu,1xu);
paths[7]:=( 1.75xu,1xu)--(4.25xu,1xu);
paths[8]:=(-0.25xu,0xu)--(2.25xu,0xu);
paths[9]:=( 2.75xu,0xu)--(4.25xu,0xu);
drawtwo paths[1];
drawtwo paths[2];
drawtwo paths[3];
drawtwo paths[4];
drawtwo paths[5];
drawone paths[6];
drawone paths[7];
drawone paths[8];
drawone paths[9];
draw (4.15xu,1.15xu)--(4.4xu,1.4xu) pensize(1pt) dashed evenly withcolor red;
arrowtwo (reverse paths[1], 0.5);
arrowtwo (reverse paths[2], 0.8);
arrowtwo (reverse paths[3], 0.5);
arrowtwo (reverse paths[4], 0.8);
arrowtwo (reverse paths[5], 0.5);
arrowone (reverse paths[6], 0.5);
arrowone (reverse paths[7], 0.5);
arrowone (reverse paths[8], 0.5);
arrowone (reverse paths[9], 0.5);
label.top(btex 1 etex, point 1 of paths[1]);
label.top(btex 1 etex, point 1 of paths[2]);
label.top(btex 1 etex, point 1 of paths[3]);
label.top(btex \textbf{1} etex, point 1 of paths[4]) withcolor 0.3[red,black];
label.top(btex 1 etex, point 1 of paths[5]);
label.bot(btex 3 etex, point 0 of paths[1]);
label.bot(btex $\mathbf{-1}$ etex, point 0 of paths[2]) withcolor 0.3[red,black];
label.bot(btex 3 etex, point 0 of paths[3]);
label.bot(btex 3 etex, point 0 of paths[4]);
label.bot(btex 3 etex, point 0 of paths[5]);
label.lft(btex 2 etex, point 0 of paths[6]);
label.lft(btex 2 etex, point 0 of paths[7]);
label.lft(btex 2 etex, point 0 of paths[8]);
label.lft(btex 2 etex, point 0 of paths[9]);
label.rt (btex 0 etex, point 1 of paths[6]);
label.rt (btex 4 etex, point 1 of paths[7]);
label.rt (btex 4 etex, point 1 of paths[8]);
label.rt (btex 4 etex, point 1 of paths[9]);
\end{mpostfig}
\caption{A pair of non-unique non-fishnet graphs at tree level.}
\label{fig:nonuniquetree}
\end{figure}

The simplest non-unique non-fishnet graphs contributing to the same bi-scalar correlator
are obtained for seven vertices at one loop and for eight vertices and tree level,
see \figref{fig:nonuniqueloop} and \figref{fig:nonuniquetree}.
These have the flavour sequences:
\<
f_{7,1} \&= (1,1,1,2,2,3,3,4,\mathbf{1},2,3,3,4,4)
\\
f_{8,0} \&= (1,1,2,\mathbf{3},4,1,1,2,2,3,3,4,\mathbf{1},2,3,3,4,4)
\>
The rule to determine a suitable sequence of evaluation parameters
directly extrapolates to these non-fishnet graphs \cite{Chicherin:2017cns,Loebbert:2025abz},
we find:
\<
s_{7,1} \&= \brk!{1,1,1,2,2,3,3,4,\mathbf{1/5},2,3,3,4,4}{}^{+4}
\\
s_{8,0} \&= \brk!{1,1,2,\mathbf{3/{-}1},0,1,1,2,2,3,3,4,\mathbf{5/1},2,3,3,4,4}{}^{+4}
\>
With these sequences,
each graph is invariant under the corresponding Yangian representation.
Due to the non-fishnet nature of these graphs, 
we do encounter instances where the number $d_j$ of adjacent propagators along the perimeter exceeds 3.
Importantly, the integer sequences of evaluation parameters for the two graphs of each pair
is not the same, it differs by $4$ units on some leg(s).
\unskip\footnote{The restricted structure of the vertices in the bi-scalar model
and the planar limit imply that the sequence of evaluation parameters modulo 4
still matches with the flavours of the corresponding legs.}
Unfortunately, this observation spoils Yangian invariance for the bi-scalar correlator
of the given flavour sequences:
Suppose the correlator $\oper{Z}$ is given as the sum of 
bi-scalar graphs $\oper{X}$ and $\oper{Y}$.
Each function $\oper{X}$ and $\oper{Y}$ is invariant
under a corresponding representation $\genyang{J}_{\oper{X},\oper{Y}}$ of the level-one Yangian generator:
\[
\genyang{J}_{\oper{X}} \oper{X} =0,
\&
\genyang{J}_{\oper{Y}} \oper{Y} =0
\]
The representations $\genyang{J}_{\oper{X}}$ and $\genyang{J}_{\oper{Y}}$ are almost the same, 
but the applicable configurations of evaluation parameters 
differ by $4$ on some site(s) $k$:
\[
\genyang{J}_{\oper{X}}-\genyang{J}_{\oper{Y}}=2\cdot4\gen{J}_k 
\]
Applying a level-one generator $\genyang{J}_{\oper{Z}}=\genyang{J}_{\oper{X}}+2\sum_j \Delta s_j\gen{J}_k$ 
with some configuration of evaluation parameters $\Delta s_j$ on top of the one for $\genyang{J}_{\oper{X}}$
yields:
\[
\genyang{J}_{\oper{Z}} \oper{Z} = 
2\sum_{j=1}^L \Delta s_j \gen{J}_j {\oper{X}}
+2\sum_{j=1}^L (\Delta s_j+4\delta_{j,k}) \gen{J}_j \oper{Y} 
\]
Owing to the inequivalent topologies of the bi-scalar graphs $\oper{X}$ and $\oper{Y}$,
there is no good reason for this combination to vanish identically
for any choice of evaluation parameter offsets $\Delta s_j$.

This finding implies that individual bi-scalar graphs
may well be invariant under Yangian generators,
but bi-scalar correlators cannot be expected to share this feature in full generality.
There is no reasonable QFT method to disentangle a correlator into its individual Feynman graph contributions
because the latter merely serve as a tool to construct correlators but have no a priori meaning on their own.
On a technical level, the failure of Yangian symmetry for generic bi-scalar correlators
might not be a concrete obstacle towards Yangian-based constructions: 
individual graph contributions may still be constructed as Yangian invariants,
and their sum will provide the appropriate correlator.
For practical purposes, most of the correlators
with reasonably simple flavour configurations
and for which a concrete computation may seem desirable,
such as fishnet graphs,
do have a unique bi-scalar graph representation.

\medskip

\begin{figure}\centering
\begin{mpostfig}
interim xu:=1.5cm;
paths[10]:=fullcircle scaled 4cm;
for j:=1 upto 8:
paths[j]:=(point (j-1.25) of paths[10]){direction (j-1.25+2) of paths[10]}
        ..{direction (j+0.25-2) of paths[10]}(point (j+0.25) of paths[10]);
endfor
drawtwo paths[1];
drawtwo paths[3];
drawtwo paths[5];
drawtwo paths[7];
drawone paths[2];
drawone paths[4];
drawone paths[6];
drawone paths[8];
draw ((0.5,0.5)--(1,1)) scaled 0.5xu shifted (paths[1] intersectionpoint paths[2]) pensize(1pt) dashed evenly withcolor red;
arrowtwo (reverse paths[1], 0.5);
arrowtwo (paths[3], 0.5);
arrowtwo (reverse paths[5], 0.5);
arrowtwo (paths[7], 0.5);
arrowone (paths[2], 0.5);
arrowone (reverse paths[4], 0.5);
arrowone (paths[6], 0.5);
arrowone (reverse paths[8], 0.5);
label.urt (btex 1 etex, point 1 of paths[1]);
label.top (btex 1 etex, point 0 of paths[3]);
label.top (btex 2 etex, point 1 of paths[2]);
label.ulft(btex 2 etex, point 0 of paths[4]);
label.ulft(btex 3 etex, point 1 of paths[3]);
label.lft (btex 3 etex, point 0 of paths[5]);
label.lft (btex 4 etex, point 1 of paths[4]);
label.llft(btex 4 etex, point 0 of paths[6]);
label.llft(btex 5 etex, point 1 of paths[5]);
label.bot (btex 5 etex, point 0 of paths[7]);
label.bot (btex 6 etex, point 1 of paths[6]);
label.lrt (btex 6 etex, point 0 of paths[8]);
label.lrt (btex 7 etex, point 1 of paths[7]);
label.rt  (btex 7 etex, point 0 of paths[1]);
label.rt  (btex 8 etex, point 1 of paths[8]);
label.urt (btex 8 etex, point 0 of paths[2]);
draw ((-0.5,-0.5)--(0.5,+0.5)) scaled 5pt pensize(2pt) withcolor 0.5[red,black];
draw ((-0.5,+0.5)--(0.5,-0.5)) scaled 5pt pensize(2pt) withcolor 0.5[red,black];
\end{mpostfig}
\caption{Octagon non-fishnet graph.}
\label{fig:escheroctagon}
\end{figure}

This is, unfortunately, not the full story:
Yangian invariance of a planar Feynman graph
requires certain constraints to hold at loop level \cite{Loebbert:2025abz}.
For the structure of the four-\hspace{0pt}dimensional bi-scalar correlators, 
these constraints require all loops to be squares.
\unskip\footnote{One can view this as a consequence of dual
conformal symmetry, where the dual graph can have only quartic vertices
which are the only conformal interactions for a regular scalar field in 4D.}
This requirement holds by construction for fishnet graphs,
but it can be violated on a non-fishnet graph with 
appropriately interconnected sheets.
\unskip\footnote{These sheets effectively introduce a conical singularity
with a surplus angle of multiples of $360^\circ$.}
A loop in a non-fishnet graph may consist of any integer multiple of 4 internal propagators,
but Yangian invariance breaks down for all higher polygons.
Here, we present a planar octagon loop graph \figref{fig:escheroctagon}.
\unskip\footnote{Graphs of this kind had been considered independently
by F.\ Loebbert and J.\ Miczajka as “amoeba” graphs
(private communication).}
The flavour sequence is given by:
\[
f_{8,1}=(1,1,2,2,3,3,4,4,1,1,2,2,3,3,4,4)
\]
The sequence of would-be evaluation parameters
as computed by the above method is:
\[
s_{8,1}=(1,1,2,2,3,3,4,4,5,5,6,6,7,7,8,8)^{+8}
\]
We observe that it displays an overall increase of $\Delta s=+8$ for a full cycle around the graph:
\[
\Delta s_{8,1}=16-\sum_{j=1}^{16}d_j=8\neq 4
\]
For proper Yangian invariance, this number ought to be $4=h^*$.
\unskip\footnote{Notably, the value of $h^*$ differs between the 
bi-scalar and brick-wall models even though they share the bi-scalar graphs.
Here, the Ward--Takahashi identities relate the bi-scalar graphs
to graphs involving fermions which explains
differences in the applicable evaluation parameter sequences.}
This mismatch is directly related to the appearance of non-square loops.
Hence, this graph is not Yangian invariant.
\unskip\footnote{It is not invariant under a standard Yangian representation
with evaluation parameters.
The dual graph certainly breaks conformal symmetry,
and along the lines of \cite{Drummond:2009fd} one should not expect
any more elaborate version of a Yangian representation
to come to the rescue of invariance.}

In conclusion, we find that only some planar correlators of the bi-scalar model are Yangian invariant. 
Others can be constructed as sums of Yangian invariants. 
Yet others cannot.

\section{Fishnet Models as a Limit}
\label{sec:limit}

Both of the fishnet models under consideration descend from 
beta/gamma-deformed $\superN=4$ supersymmetric Yang--Mills (SYM) theory 
\cite{Frolov:2005ty,Frolov:2005dj,Frolov:2005iq}
by taking particular limits 
\cite{Gurdogan:2015csr,Caetano:2016ydc}.
These “fishnet limits” combine the deformation parameters with the gauge coupling
such that only the interaction terms \eqref{eq:biscalar} and \eqref{eq:brickwall} remain.
The additional fields of $\superN=4$ SYM theory,
namely all but two complex scalars for the bi-scalar model
or an $\superN=1$ vector supermultiplet for the brick-wall model,
remain as free spectator fields:
\[@{eq:spectator}
\action_{\superN=4} \to \action_{\superN=0,1} + \action_{\text{spectator}}
\]
In the following, we will discuss whether and how the above Yangian invariances
of the action and of correlators
relate to the corresponding Yangian invariance in $\superN=4$ SYM
\cite{Beisert:2017pnr,Beisert:2018zxs,Beisert:2018ijg}.

\medskip

First, the representation of the $\alg{psu}(2,2|4)$ Yangian generators has to be twisted \cite{Garus:2017bgl,Beisert:2024wqq}. 
The generators are distinguished by their charges with respect to the deformation:
The subset of uncharged generators forms 
the residual $\alg{su}(2,2|\superN)$ Yangian symmetries of these models.
\unskip\footnote{There are some additional uncharged Cartan generators
from the internal $\alg{su}(4)$ which we will ignore.}
The representation of the level-zero generators 
changes only for some length-changing contributions.
Conversely, the bi-local representation of uncharged level-one generators
receives some contributions from pairs of oppositely charged level-zero generators.
For instance, the level-one momentum $\genyang{P}^\mu$ for beta-deformed $\superN=4$ SYM
leading to the $\superN=1$ supersymmetric brick-wall model can be sketched as:
\[
\genyang{P}_{\superN=4}
\simeq\genyang{P}_{\superN=1}+\sum_{k=1}^3\gen{Q}^k\wedge\gen{\bar Q}^k,
\&
\genyang{P}_{\superN=1}
\simeq\genyang{P}_{\superN=0}+\gen{Q}^4\wedge\gen{\bar Q}^4
\]
It has contributions of the form $\gen{Q}^k\wedge\gen{\bar Q}^k$
where the level-zero generators with $k=4$ are uncharged and belong to $\alg{su}(2,2|1)$
whereas the other supersymmetry generators with $k=1,2,3$ are charged and do not belong to $\alg{su}(2,2|1)$.
Therefore, the bi-local terms with $k=4$ are part of the plain $\alg{su}(2,2|1)$ Yangian representation,
whereas those with $k=1,2,3$ descend from the $\alg{psu}(2,2|4)$ Yangian as inhomogeneous terms.
Note that the latter terms are important to maintain Yangian symmetry in the beta-deformed model,
and they are precisely the terms which receive non-local phase dependencies.

\smallskip

Now, let us consider the limit of the invariance statement for the action.
The charged bi-local contributions to the uncharged level-one representations
are troublesome in the fishnet limit for several reasons:
First, the charged level-zero generators
do not respect the $\alg{su}(2,2|\superN)$ multiplet structure
in the sense that they can map spectator fields to physical fields of the fishnet model. 
As the spectator fields are not present in the brick-wall model action itself, 
these contributions will be missing for the purposes of overall invariance.
For example, we can consider a Yukawa term $\psi^4\psi^2\bar\phi^2$
of beta-deformed $\superN=4$ SYM
which involves the auxiliary gaugino field $\psi^4$,
and act with $\gen{Q}^1\wedge\gen{\bar Q}^1$ on it:
\[
(\gen{Q}^1\wedge\gen{\bar Q}^1) (\psi^4\psi^2\bar\phi^2)
\simeq
(\gen{\bar Q}^1\psi^4)(\gen{Q}^1\psi^2) \bar\phi^2+\ldots
\to
\partial\phi^1\.\phi^2\bar\phi^1\.\bar\phi^2+ \ldots
\]
We find a non-trivial contribution that has a similar form as
contributions from a momentum generator
acting on the term $\phi^1\phi^2\bar\phi^1\bar\phi^2$
from the brick-wall action.
Second, the representations of the level-zero generators have 
a non-trivial expansion in the deformation parameters
with some contributions suppressed in the limit.
Within the bi-local combinations, these terms
may get amplified by the divergent deformation parameters
such that they contribute non-trivially even in the fishnet limit.
In the above example, we note that the Yukawa term $\psi^4\psi^2\bar\phi^2$
is actually suppressed as $\epsilon^{+1}$ in the fishnet limit $\epsilon\to 0$.
Nevertheless, it does contribute to the Yangian representation in the limit:
\<>[rel=\simeq]
(\gen{Q}^1\wedge\gen{\bar Q}^1) (\epsilon\.\psi^4\psi^2\bar\phi^2)
\&=
\epsilon\. \partial\phi^1\.\comm{\phi^2}{\bar\phi^1}_\epsilon\.\bar\phi^2+ \ldots
\\=
\epsilon\. \partial\phi^1\.(\epsilon^{-1}\phi^2\bar\phi^1-\epsilon^{+1}\bar\phi^1\phi^2)\.\bar\phi^2+ \ldots
\\=
\frac{\epsilon}{\epsilon} \partial\phi^1\.\phi^2\bar\phi^1\.\bar\phi^2
-\epsilon^2\. \partial\phi^1\.\bar\phi^1\phi^2\.\bar\phi^2+ \ldots
\\?\to
\partial\phi^1\.\phi^2\bar\phi^1\.\bar\phi^2+ \ldots
\>
Here, the supersymmetry action $\gen{Q}^1\psi^2$
yields a twisted commutator $\comm{\phi^2}{\bar\phi^1}_\epsilon$
which has a divergent contribution that altogether survives in the limit.
More broadly, the strength of the charged bi-local terms depends non-locally on the charges
within the sequence of fields on which they act,
and it can be divergent in the limit.
Even though the limit is always perfectly finite,
the charged bi-local contributions cannot be properly constructed 
after the limit has been taken.
Nevertheless, they do have a non-trivial net contribution in the limit
which has to be effectively supplied by some inhomogeneous term in the invariance statements.
For example, acting on the unphysical sequence $\phi^3\bar\phi^3\phi^2\bar\phi^2$,
which is suppressed by $\epsilon^2$ in the fishnet limit, yields:
\[
(\gen{Q}^1\wedge\gen{\bar Q}^1) (\epsilon^2\.\phi^3\bar\phi^3\phi^2\bar\phi^2)
\simeq
\frac{\epsilon^2}{\epsilon^2} (\gen{\bar Q}^1\phi^3)\bar\phi^3\phi^2(\gen{Q}^1\bar\phi^2)
\to
\bar\psi^2\.\bar\phi^3\phi^2\.\psi^3+ \ldots
\]
Here, the divergent factor $\epsilon^{-2}$
arises from the particular charge configuration in the bi-local variation.
Note that the form of this term 
compares well to a contribution of $\gen{Q}^4\wedge\gen{\bar Q}^4$
acting on the term $\bar\phi^2\.\bar\phi^3\phi^2\.\phi^3$
from the brick-wall action.
Third, the local representation of the level-one generators
should be adjusted because it incorporates
the short-distance regularisation for the replacement bi-local terms.

Unfortunately, the effective and/or inhomogeneous terms
for the invariance statement depend on the original action. 
As such, we cannot universally derive the proper reduction of the Yangian level-one representation;
we can merely propose a concrete form applicable for the purposes of invariance of the action.
Nevertheless, we can draw inspiration from a similar case, namely the reduction of the Yangian algebra 
for the scattering picture within the coordinate Bethe ansatz.
There, one finds that the effect of additional bi-local terms
in the original Yangian algebra
is summarised by adjustments of the evaluation parameters
in the applicable Yangian subalgebra \cite[chapter 7]{BeisertIntegrability:2016}.
We observe that the present case is compatible with this kind of reduction.

\smallskip

Let us finally turn towards the Yangian invariance for correlators.
We have seen in \secref{sec:correlators} that some correlators
satisfy an invariance statement with properly chosen evaluation parameters for the external legs
while the evaluation parameters for other correlators
cannot be chosen consistently to ensure Yangian invariance.
Now, the above discussion on taking the limit for invariance of the action
equivalently applies to correlation functions:
Correlators of the fishnet models can be derived
as the fishnet limit of the corresponding correlators in beta/gamma-deformed $\superN=4$ SYM.
Nevertheless, establishing their Yangian invariance
requires not only the leading-order correlator but also sub-leading contributions
which are amplified by divergent terms in the symmetry representation.
As the latter are not expressible in the limiting fishnet models alone, 
they would have to be replaced by some effective terms,
and hence Yangian invariance in the limit cannot be taken for granted.
The contributions from evaluation parameters serve this purpose,
and they work well for correlators which are represented by a unique graph
whose loops are all squares.
However, not all correlators are of this type
and invariance cannot be achieved using evaluation parameters.
It is conceivable, yet not likely,
that some other prescription for effective terms
might restore Yangian invariance for these correlators.

\medskip

Altogether, we make the following observations
concerning the limiting Yangian representations:
The representation of the uncharged generators
has a finite limit which agrees with the
expressions in the plain $\alg{su}(2,2|\superN)$ Yangian
\[
\genyang{J}_{\superN=4} \to \genyang{J}_{\superN=0,1}
\&
\genyang{J}_{\superN=0,1;L,k} = c_{L,k}\gen{J}_k
\]
In the limit, all charged bi-local contributions
as well as all local contributions to level-one generators are dropped
and effectively replaced by evaluation parameters $c_{L,k}$.
The action of the $\superN=1$ brick-wall model becomes invariant 
with a non-trivial configuration of evaluation parameters $c_{L,k}$
acting on the quartic scalar term \eqref{eq:anomsusy}.
The action of the bi-scalar model requires no effective replacement terms, $c_{L,k}=0$.
For correlation functions, the pattern of evaluation parameters
depends on the configuration of the external field flavours;
some correlators are invariant with a suitable choice of evaluation parameters
while for others there are no consistent choices.

\section{Conclusions}
\label{sec:conclusions}

We have shown that the equations of motion and the action 
in the bi-scalar and brick-wall models
\eqref{eq:biscalar} and \eqref{eq:brickwall}
are classically invariant under an $\alg{su}(2,2|\superN)$ Yangian symmetry
in the planar limit.
The appropriate representation to achieve invariance 
involves some non-trivial assignments of evaluation parameters
which are correlated to the non-vanishing of the dual Coxeter numbers
$h^*=4-\superN$ for the conformal symmetry algebras in these models.

We have further discussed particular bi-scalar correlators
where Yangian invariance breaks.
The existence of these correlators invalidates
two common assumptions about planar correlators in these models: 
Firstly, a bi-scalar correlator need not be represented by a single Feynman graph alone,
it may be a sum of two or more Feynman graphs with inequivalent topologies.
Secondly, a loop in a bi-scalar correlator need not necessarily have four sides,
it may have any multiple of four sides. 
We have then argued that Yangian invariance fails for particular bi-scalar correlators,
which serve as concrete counter-examples
to the broad exceptional classes of graphs where invariance has previously been established.
\unskip\footnote{We expect that for more elaborate flavour configurations of the external legs, 
failure of Yangian invariance will be the rule rather than an exception.}
Hence, Yangian symmetry does not apply in generality in these models.
By analogous reasoning, we expect these findings to also apply to the brick-wall model.
\unskip\footnote{In particular, there can be internal loops of non-minimal size,
which are detrimental to Yangian invariance.}
One might interpret the failure of Yangian invariance for quantum correlators
as an anomaly:
the rules by which vertices of the classical action are assembled
into Feynman graphs, 
are structurally incompatible with the bi-local workings
of level-one Yangian generators.
Altogether, the Yangian algebra fails to be a symmetry of the full quantum planar model.

Finally, we have elaborated on the connection to $\superN=4$ SYM and its beta/gamma-\hspace{0pt}deformation.
To some extent, Yangian symmetry derives from Yangian symmetry in the $\superN=4$ SYM ancestor.
However, the limit required to obtain the fishnet models is singular,
and this singularity obscures Yangian symmetry.
We effectively restored Yangian symmetry
for the equations of motion and the action
by means of suitably chosen patterns of evaluation parameters,
which in the case of the brick-wall model are applicable to specific terms.
However, in correlators these evaluation parameters
cannot always be assigned consistently, and therefore Yangian symmetry fails.
\unskip\footnote{We do not exclude the remote possibility
of using effective terms generalising evaluation parameters
in order to restore invariance.
Such terms would be constrained by having
to respect the representation property of the Yangian algebra.}
Does this failure have implications for Yangian invariance of correlators in
$\superN=4$ SYM ancestor?
Not necessarily. 
The failure is based on inconsistent assignments of the evaluation parameters.
The evaluation parameter mismatches all turned out to be proportional
to the dual Coxeter number $h^*=4-\superN$.
For the $\superN=4$ superconformal Yangian
the applicable dual Coxeter number is zero.
Our findings imply no particular restriction
for Yangian invariance of planar correlators in $\superN=4$ SYM.
Indeed, several planar correlators in $\superN=4$ SYM have been
demonstrated to be properly Yangian invariant (up to regularisation) \cite{Beisert:2018ijg};
none of these involved evaluation parameters.

In conclusion, our results suggest that a non-zero dual Coxeter number $h^*$
and/or the requirement of using evaluation parameters
represents an obstruction towards the full applicability of quantum Yangian symmetry in a planar QFT model.
Even though the four-dimensional fishnet models appeared to evade this intuition,
ultimately, they are no counter-examples to the obstruction.
This leaves $\superN=4$ SYM as well as its deformations and orbifolds
as the only four-dimensional planar gauge theory models 
with (potentially) full Yangian symmetry at the quantum level.
Similar conclusions will apply to other spacetime dimensions, 
e.g.\ in three dimensions, $\superN=6$ supersymmetric Chern--Simons models
will play an analogously distinguished role concerning full Yangian symmetry
as $\superN=4$ SYM in four dimensions.

\medskip

Some points may deserve further clarifications:

We have discarded the double-trace contributions to the action, 
which have some relevance in the planar limit at loop order, see \cite{Sieg:2016vap,Grabner:2017pgm}. 
It appears unlikely that such terms contribute to the correlators
discussed here, or if they do, that they could change our conclusions.
A more careful analysis would be needed to settle this question.

Similarly, we have based our considerations
on the standard class of tensor products of Yangian evaluation representations
which comes to use in the context of (inhomogeneous) spin chain models.
We cannot say whether or not some other type of (consistent) Yangian representation
on bi-scalar graphs and/or correlators could potentially resolve the above issues.

The failure of full Yangian symmetry
suggests that integrability also breaks for planar fishnet models.
For instance, one may argue that the conserved charges defining integrability
are typically obtained as elements of the centre of Yangian,
and it appears unlikely that the issues plaguing Yangian symmetry
will not extend to its centre.
However, a proper general definition of integrability for the class of planar QFT models is lacking,
and the question of integrability for fishnet models evidently depends on it.
We thus cannot strictly rule out integrability for these models,
and in any case, methods of integrability have already proved useful for selected observables and processes.

\begin{figure}\centering
\begin{mpostfig}[opt=align]
interim xu:=1.5cm;
paths[11]:=fullcircle scaled 1cm shifted (-2cm,0);
paths[12]:=fullcircle rotated 180 scaled 1cm shifted (+2cm,0);
fill paths[11] withcolor 0.8white;
fill paths[12] withcolor 0.8white;
draw subpath (-2,2) of paths[11];
draw subpath (-2,2) of paths[12];
draw subpath (2,6) of paths[11] dashed evenly;
draw subpath (2,6) of paths[12] dashed evenly;
paths[1]:=(point -0.75 of paths[12]){dir +146.25}..{dir -101.25}(point +1.75 of paths[11]);
paths[2]:=(point -1.75 of paths[12]){dir +101.25}..{dir -146.25}(point +0.75 of paths[11]);
paths[3]:=(point +1.25 of paths[11]){dir  +56.25}..{dir +168.75}(point -0.25 of paths[11]);
paths[4]:=(point +0.25 of paths[11]){dir  +11.25}..{dir +123.75}(point -1.25 of paths[11]);
paths[5]:=(point -0.75 of paths[11]){dir  -33.75}..{dir  +78.75}(point +1.75 of paths[12]);
paths[6]:=(point -1.75 of paths[11]){dir  -78.75}..{dir  +33.75}(point +0.75 of paths[12]);
paths[7]:=(point +1.25 of paths[12]){dir -123.75}..{dir  -11.25}(point -0.25 of paths[12]);
paths[8]:=(point +0.25 of paths[12]){dir -168.75}..{dir  -56.25}(point -1.25 of paths[12]);
drawtwo paths[1];
drawtwo paths[3];
drawtwo paths[5];
drawtwo paths[7];
drawone paths[2];
drawone paths[4];
drawone paths[6];
drawone paths[8];
arrowtwo (reverse paths[1], 0.8);
arrowtwo (paths[3], 0.6);
arrowtwo (reverse paths[5], 0.8);
arrowtwo (paths[7], 0.6);
arrowone (paths[2], 0.8);
arrowone (reverse paths[4], 0.6);
arrowone (paths[6], 0.8);
arrowone (reverse paths[8], 0.6);
draw ((-0.5,-0.5)--(0.5,+0.5)) scaled 5pt pensize(2pt) withcolor 0.5[red,black];
draw ((-0.5,+0.5)--(0.5,-0.5)) scaled 5pt pensize(2pt) withcolor 0.5[red,black];
\end{mpostfig}
\qquad
\begin{mpostfig}[opt=align]
interim xu:=1.5cm;
paths[11]:=fullcircle scaled 1cm shifted (-1cm,+1cm);
paths[12]:=fullcircle scaled 1cm shifted (+1cm,+1cm);
paths[13]:=fullcircle scaled 1cm shifted (+1cm,-1cm);
paths[14]:=fullcircle scaled 1cm shifted (-1cm,-1cm);
fill paths[11] withcolor 0.8white;
fill paths[12] withcolor 0.8white;
fill paths[13] withcolor 0.8white;
fill paths[14] withcolor 0.8white;
draw subpath (-2,0) of paths[11];
draw subpath (-4,-2) of paths[12];
draw subpath (2,4) of paths[13];
draw subpath (0,2) of paths[14];
draw subpath (0,6) of paths[11] dashed evenly;
draw subpath (-2,4) of paths[12] dashed evenly;
draw subpath (-4,2) of paths[13] dashed evenly;
draw subpath (2,8) of paths[14] dashed evenly;
paths[1]:=(point -2.75 of paths[12]){dir -123.75}..{dir +168.75}(point -0.25 of paths[11]);
paths[2]:=(point -3.75 of paths[12]){dir -168.75}..{dir +123.75}(point -1.25 of paths[11]);
paths[3]:=(point -0.75 of paths[11]){dir  -33.75}..{dir -101.25}(point +1.75 of paths[14]);
paths[4]:=(point -1.75 of paths[11]){dir  -78.75}..{dir -146.25}(point +0.75 of paths[14]);
paths[5]:=(point +1.25 of paths[14]){dir  +56.25}..{dir  -11.25}(point +3.75 of paths[13]);
paths[6]:=(point +0.25 of paths[14]){dir  +11.25}..{dir  -56.25}(point +2.75 of paths[13]);
paths[7]:=(point +3.25 of paths[13]){dir +146.25}..{dir  +78.75}(point -2.25 of paths[12]);
paths[8]:=(point +2.25 of paths[13]){dir +101.25}..{dir  +33.75}(point -3.25 of paths[12]);
drawtwo paths[1];
drawtwo paths[3];
drawtwo paths[5];
drawtwo paths[7];
drawone paths[2];
drawone paths[4];
drawone paths[6];
drawone paths[8];
arrowtwo (reverse paths[1], 0.55);
arrowtwo (paths[3], 0.5);
arrowtwo (reverse paths[5], 0.55);
arrowtwo (paths[7], 0.5);
arrowone (paths[2], 0.55);
arrowone (reverse paths[4], 0.5);
arrowone (paths[6], 0.55);
arrowone (reverse paths[8], 0.5);
draw ((-0.5,-0.5)--(0.5,+0.5)) scaled 5pt pensize(2pt) withcolor 0.5[red,black];
draw ((-0.5,+0.5)--(0.5,-0.5)) scaled 5pt pensize(2pt) withcolor 0.5[red,black];
\end{mpostfig}
\caption{Octagon loops in two- and four-point functions of local operators.}
\label{fig:octagondimension}
\end{figure}

It even remains conceivable that non-invariant bi-scalar graphs containing internal non-square loops
are nonetheless constructible using other worldsheet patching techniques of integrability.
For instance, it would be worthwhile to figure out whether the bi-scalar graph in
\figref{fig:escheroctagon} with an octagon loop could somehow be constructed
using the hexagon approach to planar correlators in integrable QFT models \cite{Basso:2015zoa,Basso:2018cvy,Olivucci:2021cfy}.
It will be equally relevant to consider planar octagon loop contributions
to other observables that can be addressed by methods of integrability.
In particular, octagon loops may contribute to the planar anomalous dimensions
of local operators composed from at least two copies of all four flavours of scalar fields
($\sim\tr\phi_1^{n_1}\phi_2^{n_2}\bar\phi_1^{\bar n_1}\bar\phi_2^{\bar n_2}$
with $n_1,n_2,\bar n_1,\bar n_2\geq 2$),
at eighth perturbative order or beyond,
see \figref{fig:octagondimension}.
\unskip\footnote{The fact that 8 fields are connected by merely 4 intermediate propagators
may seem intriguing, but apparently does not preclude a contribution,
see e.g.\ \cite{Gromov:2019aku,Gromov:2019bsj,Gromov:2019jfh},
even to the anomalous dimensions.}
\footnote{Selected local operators in non-unitary Fishnet models receive Jordan block contributions
to the planar anomalous dimension matrix which lead to additive scaling behaviour \cite[section 7.2]{Gromov:2017cja}.
While this is a peculiar effect, we are not aware that it relates to a breakdown of integrability;
we rather consider it independent of our considerations on the breakdown of Yangian symmetry.}
Likewise, four-point correlators of local operators with at least one copy
of all four flavours of scalar fields receive such contributions.
The established results on Yangian invariance admittedly make no statement
on the integrability for correlators of coincident external fields;
nevertheless, it is very suggestive that integrability will be infringed
by the presence of an internal octagon loop
if Yangian symmetry gets broken by precisely this configuration
in another case.
Hence, does the quantum spectral curve, see \cite{Kazakov:2018ugh},
for the bi-scalar model actually predict the planar anomalous dimensions
of all local operators correctly at all perturbative orders?
How do octagon loop contributions fit in
with the spectral curve and with the holographic dual model 
proposed in \cite{Gromov:2019aku,Gromov:2019bsj,Gromov:2019jfh}?

\pdfbookmark[1]{Acknowledgements}{ack}
\section*{Acknowledgements}

We are grateful to F.~Loebbert for extensive discussion on uniqueness
and Yangian invariance of fishnet diagrams.
We also thank him and V.~Kazakov for comments and discussions on the draft.
The work of NB is partially supported by the Swiss National Science Foundation
through the NCCR SwissMAP.
This work is based in parts on results from BK's master thesis at ETH Zürich.

\appendix

\section{Uniqueness of Graphs}
\label{sec:uniqueness}

In this appendix, we comment on the uniqueness of bi-scalar graphs
contributing to bi-scalar correlators.
In particular, we would like to address two questions:
Is there a unique bi-scalar graph
for a given sequence of $E$ flavours $f_k$ of the external legs?
And is there a unique bi-scalar graph for a given perimeter configuration
in terms of the number of consecutive propagators $d_k$ along the perimeter
(together with the starting flavour $f_1$)?
Alternatively, we may describe the latter data in terms of a 
sequence of evaluation parameters $s_k$
(together with the overall evaluation parameter shift $\Delta s$):
\[@{eq:svsd}
s_k=f_1+\sum_{j=1}^{k-1}(1-d_j),
\&
\Delta s=E-\sum_{j=1}^{E}d_j
\]

\medskip

Before doing so, let us state some useful identities
for the characteristic numbers of bi-scalar graphs
which are obtained by standard considerations of Feynman graphs:
The sequence of flavours immediately determines
the number of legs of types $\phi^{1,2}$ as
\[
E_{1,2}=\frac{E}{4} \mp \sum_{k=1}^E \frac{(-1)^{f_j}}{4}
\]
Note that the number of conjugate fields $\bar\phi^{1,2}$
must match due to charge conservation. 
In total we have $E=2E_1+2E_2$ legs.
Given the perimeter configuration,
we can deduce the number of propagators of each flavour along the perimeter as
\[
P_{1,2}
=\sum_{j=1}^{E}\frac{d_j}{2} \pm \sum_{k=1}^E \frac{(-1)^{f_j}}{2}
=\frac{E}{2}-\frac{\Delta s}{2} \pm \sum_{k=1}^E \frac{(-1)^{s_j}}{2}
\]
This follows from the fact that the flavours of perimeter propagators must alternate.
The total number of propagators along the perimeter reads
\[
P=\sum_{j=1}^{E}d_j=E-\Delta s
\]

Importantly, we can determine the number of vertices $V$ abstractly as:
\[
V=\sum_{j<k=1}^E \sin\brk!{\half\pi (s_k-s_j)}
\]
Note that this formula computes the fishnet area enclosed by the fishnet shape
which equivalently describes the total twist of the flavour configuration 
when viewed as a deformation of $\superN=4$ SYM.
Given $V$ and the structure of vertices, we can deduce further useful characteristic numbers.
The numbers of internal propagators of each flavour reads:
\[
I_{1,2}=V-E_{1,2}
\]
Euler's formula also determines the number of loops $L$ as:
\[
L=V+1-\half E
\]
The sum of loop valences $N_j$, i.e.\ the number of propagators around each loop $j$, 
is described by $\Delta s$ as:
\[
\sum_{j=1}^L N_j = 4L+\Delta s = 4L+E-\sum_{j=1}^{E}d_j
\]

\medskip

\begin{figure}\centering
\begin{mpostfig}
interim xu:=1.5cm;
paths[10]:=fullcircle scaled 4cm;
for j:=1 upto 8:
paths[j]:=(point (j-1.75) of paths[10]){direction (j-1.75+2) of paths[10]}
        ..{direction (j-0.25-2) of paths[10]}(point (j-0.25) of paths[10]);
endfor
for j:=1 upto 4:
paths[10+j]:=(point (2j-2) of paths[10]){direction (2j-2+2) of paths[10]}
        ..(point (2j-1) of paths[10] scaled 0.37)
        ..{direction (2j-2) of paths[10]}(point (2j) of paths[10]);
endfor
drawtwo paths[1];
drawtwo paths[3];
drawtwo paths[5];
drawtwo paths[7];
drawone paths[2];
drawone paths[4];
drawone paths[6];
drawone paths[8];
drawone paths[11];
drawone paths[13];
draw ((0.5,0)--(1,0)) scaled 0.7xu rotated 22.5 shifted (paths[1] intersectionpoint paths[2]) pensize(1pt) dashed evenly withcolor red;
arrowtwo (reverse paths[1], 0.56);
arrowtwo (paths[3], 0.56);
arrowtwo (reverse paths[5], 0.56);
arrowtwo (paths[7], 0.56);
arrowone (paths[2], 0.5);
arrowone (reverse paths[4], 0.5);
arrowone (paths[6], 0.5);
arrowone (reverse paths[8], 0.5);
arrowone (paths[11], 1);
arrowone (paths[13], 1);
label.urt (btex 1 etex, point 1 of paths[1]);
label.urt (btex 1 etex, point 0 of paths[3]);
label.top (btex 2 etex, point 1 of paths[2]);
label.top (btex 2 etex, point 2 of paths[11]);
label.top (btex 2 etex, point 0 of paths[4]);
label.ulft(btex 3 etex, point 1 of paths[3]);
label.ulft(btex 3 etex, point 0 of paths[5]);
label.lft (btex 4 etex, point 1 of paths[4]);
label.lft (btex 4 etex, point 0 of paths[13]);
label.lft (btex 4 etex, point 0 of paths[6]);
label.llft(btex 5 etex, point 1 of paths[5]);
label.llft(btex 5 etex, point 0 of paths[7]);
label.bot (btex 6 etex, point 1 of paths[6]);
label.bot (btex 6 etex, point 2 of paths[13]);
label.bot (btex 6 etex, point 0 of paths[8]);
label.lrt (btex 7 etex, point 1 of paths[7]);
label.lrt (btex 7 etex, point 0 of paths[1]);
label.rt  (btex 8 etex, point 1 of paths[8]);
label.rt  (btex 8 etex, point 0 of paths[11]);
label.rt  (btex 8 etex, point 0 of paths[2]);
draw ((-0.5,-0.5)--(0.5,+0.5)) scaled 5pt pensize(2pt) withcolor 0.5[red,black];
draw ((-0.5,+0.5)--(0.5,-0.5)) scaled 5pt pensize(2pt) withcolor 0.5[red,black];
\end{mpostfig}
\qquad
\begin{mpostfig}
interim xu:=1.5cm;
paths[10]:=fullcircle scaled 4cm;
for j:=1 upto 8:
paths[j]:=(point (j-1.75) of paths[10]){direction (j-1.75+2) of paths[10]}
        ..{direction (j-0.25-2) of paths[10]}(point (j-0.25) of paths[10]);
endfor
for j:=1 upto 4:
paths[10+j]:=(point (2j-2) of paths[10]){direction (2j-2+2) of paths[10]}
        ..(point (2j-1) of paths[10] scaled 0.37)
        ..{direction (2j-2) of paths[10]}(point (2j) of paths[10]);
endfor
drawtwo paths[1];
drawtwo paths[3];
drawtwo paths[5];
drawtwo paths[7];
drawone paths[2];
drawone paths[4];
drawone paths[6];
drawone paths[8];
drawone paths[12];
drawone paths[14];
draw ((0.5,0)--(1,0)) scaled 0.7xu rotated 22.5 shifted (paths[1] intersectionpoint paths[2]) pensize(1pt) dashed evenly withcolor red;
arrowtwo (reverse paths[1], 0.44);
arrowtwo (paths[3], 0.44);
arrowtwo (reverse paths[5], 0.44);
arrowtwo (paths[7], 0.44);
arrowone (paths[2], 0.5);
arrowone (reverse paths[4], 0.5);
arrowone (paths[6], 0.5);
arrowone (reverse paths[8], 0.5);
arrowone (reverse paths[12], 1);
arrowone (reverse paths[14], 1);
label.urt (btex 1 etex, point 1 of paths[1]);
label.urt (btex 1 etex, point 0 of paths[3]);
label.top (btex 2 etex, point 1 of paths[2]);
label.top (btex 2 etex, point 2 of paths[11]);
label.top (btex 2 etex, point 0 of paths[4]);
label.ulft(btex 3 etex, point 1 of paths[3]);
label.ulft(btex 3 etex, point 0 of paths[5]);
label.lft (btex 4 etex, point 1 of paths[4]);
label.lft (btex 4 etex, point 0 of paths[13]);
label.lft (btex 4 etex, point 0 of paths[6]);
label.llft(btex 5 etex, point 1 of paths[5]);
label.llft(btex 5 etex, point 0 of paths[7]);
label.bot (btex 6 etex, point 1 of paths[6]);
label.bot (btex 6 etex, point 2 of paths[13]);
label.bot (btex 6 etex, point 0 of paths[8]);
label.lrt (btex 7 etex, point 1 of paths[7]);
label.lrt (btex 7 etex, point 0 of paths[1]);
label.rt  (btex 8 etex, point 1 of paths[8]);
label.rt  (btex 8 etex, point 0 of paths[11]);
label.rt  (btex 8 etex, point 0 of paths[2]);
draw ((-0.5,-0.5)--(0.5,+0.5)) scaled 5pt pensize(2pt) withcolor 0.5[red,black];
draw ((-0.5,+0.5)--(0.5,-0.5)) scaled 5pt pensize(2pt) withcolor 0.5[red,black];
\end{mpostfig}
\caption{Non-fishnet graphs with equal perimeter and different interiors.}
\label{fig:octagonflip}
\end{figure}

Next, let us discuss some further non-unique bi-scalar graphs
next to the examples already given in \secref{sec:correlators}. 
We will use these to restrain the maximum applicability
of potential uniqueness statements.

We have already shown in \figref{fig:nonuniqueloop} and \figref{fig:nonuniquetree}
that there can be inequivalent bi-scalar graphs
for a given sequence of flavours $f_k$.
The pair of graphs in \figref{fig:octagonflip} demonstrates
that there can be two inequivalent
bi-scalar graphs for a given perimeter configuration:
\[
s_{12,3}=(1,1,2,2,2,3,3,4,4,4,5,5,6,6,6,7,7,8,8,8)^{+\mathbf{8}}
\]
We note that the two bi-scalar graphs are related by a simple 
reconfiguration of a pair of internal lines facing in opposite directions.
Such a move can be applied to wide classes of graphs
provided that their interior is sufficiently complex.
Hence there will be many further examples of non-unique graphs
with the same perimeter configuration.
Let us make two further related comments before moving on
to refinements of the two questions.

\begin{figure}\centering
\begin{mpostfig}
interim xu:=1.5cm;
paths[10]:=fullcircle scaled 4cm;
for j:=1 upto 8:
paths[j]:=(point (j-1.75) of paths[10]){direction (j-1.75+2) of paths[10]}
        ..{direction (j-0.25-2) of paths[10]}(point (j-0.25) of paths[10]);
endfor
for j:=1 upto 4:
paths[10+j]:=(point (2j-2) of paths[10]){direction (2j-2+2) of paths[10]}
        ..(point (2j-1) of paths[10] scaled 0.37)
        ..{direction (2j-2) of paths[10]}(point (2j) of paths[10]);
endfor
drawtwo paths[1];
drawtwo paths[3];
drawtwo paths[5];
drawone paths[2];
drawone paths[4];
draw paths[11] pensize(2.25pt) dashed evenly withcolor colone;
draw paths[13] pensize(2.25pt) dashed evenly withcolor colone;
draw ((0.5,0)--(1,0)) scaled 0.7xu rotated 22.5 shifted (paths[1] intersectionpoint paths[2]) pensize(1pt) dashed evenly withcolor red;
arrowtwo (reverse paths[1], 0.56);
arrowtwo (paths[3], 0.56);
arrowtwo (reverse paths[5], 0.56);
arrowone (paths[2], 0.5);
arrowone (reverse paths[4], 0.5);
arrowone (paths[11], 1);
arrowone (paths[13], 1);
label.urt (btex 1 etex, point 1 of paths[1]);
label.urt (btex 1 etex, point 0 of paths[3]);
label.top (btex 2 etex, point 1 of paths[2]);
label.top (btex 2 etex, point 2 of paths[11]);
label.top (btex 2 etex, point 0 of paths[4]);
label.ulft(btex 3 etex, point 1 of paths[3]);
label.ulft(btex 3 etex, point 0 of paths[5]);
label.lft (btex 4 etex, point 1 of paths[4]);
label.lft (btex 4 etex, point 0 of paths[13]);
label.llft(btex 5 etex, point 1 of paths[5]);
label.bot (btex 6 etex, point 2 of paths[13]) withcolor 0.6red;
label.lrt (btex 3 etex, point 0 of paths[1]);
label.rt  (btex 4 etex, point 0 of paths[11]);
label.rt  (btex 4 etex, point 0 of paths[2]);
\end{mpostfig}
\qquad
\begin{mpostfig}
interim xu:=1.5cm;
paths[10]:=fullcircle scaled 4cm;
for j:=1 upto 8:
paths[j]:=(point (j-1.75) of paths[10]){direction (j-1.75+2) of paths[10]}
        ..{direction (j-0.25-2) of paths[10]}(point (j-0.25) of paths[10]);
endfor
for j:=1 upto 4:
paths[10+j]:=(point (2j-2) of paths[10]){direction (2j-2+2) of paths[10]}
        ..(point (2j-1) of paths[10] scaled 0.37)
        ..{direction (2j-2) of paths[10]}(point (2j) of paths[10]);
endfor
drawtwo paths[1];
drawtwo paths[3];
drawtwo paths[5];
drawone paths[2];
drawone paths[4];
draw paths[12] pensize(2.25pt) dashed evenly withcolor colone;
draw paths[14] pensize(2.25pt) dashed evenly withcolor colone;
draw ((0.5,0)--(1,0)) scaled 0.7xu rotated 22.5 shifted (paths[1] intersectionpoint paths[2]) pensize(1pt) dashed evenly withcolor red;
arrowtwo (reverse paths[1], 0.44);
arrowtwo (paths[3], 0.44);
arrowtwo (reverse paths[5], 0.44);
arrowone (paths[2], 0.5);
arrowone (reverse paths[4], 0.5);
arrowone (reverse paths[12], 1);
arrowone (reverse paths[14], 1);
label.urt (btex 1 etex, point 1 of paths[1]);
label.urt (btex 1 etex, point 0 of paths[3]);
label.top (btex 2 etex, point 1 of paths[2]);
label.top (btex 2 etex, point 2 of paths[11]);
label.top (btex 2 etex, point 0 of paths[4]);
label.ulft(btex 3 etex, point 1 of paths[3]);
label.ulft(btex 3 etex, point 0 of paths[5]);
label.lft (btex 4 etex, point 1 of paths[4]);
label.lft (btex 4 etex, point 0 of paths[13]);
label.llft(btex 5 etex, point 1 of paths[5]);
label.bot (btex 2 etex, point 2 of paths[13]) withcolor 0.6red;
label.lrt (btex 3 etex, point 0 of paths[1]);
label.rt  (btex 4 etex, point 0 of paths[11]);
label.rt  (btex 4 etex, point 0 of paths[2]);
\end{mpostfig}
\caption{Basic reconfiguration move and corresponding adjustment to evaluation parameter sequence.}
\label{fig:basicflip}
\end{figure}

We can reduce the reconfiguration move to its essentials
while maintaining the connectedness of the graph, and obtain the 
pair of graphs in \figref{fig:basicflip}.
This reproduces precisely the simplest pair of inequivalent graphs with identical flavour configurations
in \figref{fig:nonuniqueloop}.
Note that the pair of graphs shown in \figref{fig:nonuniquetree} can be related
by a sequence of two such moves (where connectedness is dropped in the intermediate stage).

\begin{figure}\centering
\begin{mpostfig}
interim xu:=1.5cm;
paths[10]:=fullcircle scaled 4cm;
for j:=1 upto 8:
paths[j]:=(point (j-1.75) of paths[10]){direction (j-1.75+2) of paths[10]}
        ..{direction (j-0.25-2) of paths[10]}(point (j-0.25) of paths[10]);
endfor
for j:=1 upto 4:
paths[10+j]:=(point (2j-2) of paths[10]){direction (2j-2+2) of paths[10]}
        ..(point (2j-1) of paths[10] scaled 0.37)
        ..{direction (2j-2) of paths[10]}(point (2j) of paths[10]);
endfor
drawtwo paths[1];
drawtwo paths[3];
drawtwo paths[5];
drawtwo paths[7];
drawone paths[4];
drawone paths[6];
drawone paths[8];
drawone paths[11];
drawone paths[13];
draw ((0.7,0)--(1.2,0)) scaled 2cm rotated 45 pensize(1pt) dashed evenly withcolor red;
arrowtwo (reverse paths[1], 0.56);
arrowtwo (paths[3], 0.56);
arrowtwo (reverse paths[5], 0.56);
arrowtwo (paths[7], 0.56);
arrowone (reverse paths[4], 0.5);
arrowone (paths[6], 0.5);
arrowone (reverse paths[8], 0.5);
arrowone (paths[11], 1);
arrowone (paths[13], 1);
label.urt (btex 9 etex, point 1 of paths[1]);
label.urt (btex 1 etex, point 0 of paths[3]);
label.top (btex 2 etex, point 2 of paths[11]);
label.top (btex 2 etex, point 0 of paths[4]);
label.ulft(btex 3 etex, point 1 of paths[3]);
label.ulft(btex 3 etex, point 0 of paths[5]);
label.lft (btex 4 etex, point 1 of paths[4]);
label.lft (btex 4 etex, point 0 of paths[13]);
label.lft (btex 4 etex, point 0 of paths[6]);
label.llft(btex 5 etex, point 1 of paths[5]);
label.llft(btex 5 etex, point 0 of paths[7]);
label.bot (btex 6 etex, point 1 of paths[6]);
label.bot (btex 6 etex, point 2 of paths[13]);
label.bot (btex 6 etex, point 0 of paths[8]);
label.lrt (btex 7 etex, point 1 of paths[7]);
label.lrt (btex 7 etex, point 0 of paths[1]);
label.rt  (btex 8 etex, point 1 of paths[8]);
label.rt  (btex 8 etex, point 0 of paths[11]);
draw ((-0.5,-0.5)--(0.5,+0.5)) scaled 5pt pensize(2pt) withcolor 0.5[red,black];
draw ((-0.5,+0.5)--(0.5,-0.5)) scaled 5pt pensize(2pt) withcolor 0.5[red,black];
\end{mpostfig}
\qquad
\begin{mpostfig}
interim xu:=1.5cm;
paths[10]:=fullcircle scaled 4cm;
for j:=1 upto 8:
paths[j]:=(point (j-1.75) of paths[10]){direction (j-1.75+2) of paths[10]}
        ..{direction (j-0.25-2) of paths[10]}(point (j-0.25) of paths[10]);
endfor
for j:=1 upto 4:
paths[10+j]:=(point (2j-2) of paths[10]){direction (2j-2+2) of paths[10]}
        ..(point (2j-1) of paths[10] scaled 0.37)
        ..{direction (2j-2) of paths[10]}(point (2j) of paths[10]);
endfor
drawtwo paths[1];
drawtwo paths[3];
drawtwo paths[5];
drawtwo paths[7];
drawone paths[4];
drawone paths[6];
drawone paths[8];
drawone paths[12];
drawone paths[14];
draw ((0.7,0)--(1.2,0)) scaled 2cm rotated 45 pensize(1pt) dashed evenly withcolor red;
arrowtwo (reverse paths[1], 0.44);
arrowtwo (paths[3], 0.44);
arrowtwo (reverse paths[5], 0.44);
arrowtwo (paths[7], 0.44);
arrowone (reverse paths[4], 0.5);
arrowone (paths[6], 0.5);
arrowone (reverse paths[8], 0.5);
arrowone (reverse paths[12], 1);
arrowone (reverse paths[14], 1);
label.urt (btex 9 etex, point 1 of paths[1]);
label.urt (btex 1 etex, point 0 of paths[3]);
label.top (btex 2 etex, point 2 of paths[11]);
label.top (btex 2 etex, point 0 of paths[4]);
label.ulft(btex 3 etex, point 1 of paths[3]);
label.ulft(btex 3 etex, point 0 of paths[5]);
label.lft (btex 4 etex, point 1 of paths[4]);
label.lft (btex 4 etex, point 0 of paths[13]);
label.lft (btex 4 etex, point 0 of paths[6]);
label.llft(btex 5 etex, point 1 of paths[5]);
label.llft(btex 5 etex, point 0 of paths[7]);
label.bot (btex 6 etex, point 1 of paths[6]);
label.bot (btex 6 etex, point 2 of paths[13]);
label.bot (btex 6 etex, point 0 of paths[8]);
label.lrt (btex 7 etex, point 1 of paths[7]);
label.lrt (btex 7 etex, point 0 of paths[1]);
label.rt  (btex 8 etex, point 1 of paths[8]);
label.rt  (btex 8 etex, point 0 of paths[11]);
\end{mpostfig}
\caption{Two graphs with equal evaluation parameter sequences (up to periodicity),
but inequivalent interiors.}
\label{fig:inequivalent}
\end{figure}

In \figref{fig:inequivalent} we provide two bi-scalar graphs
with identical sequences of evaluation parameters $s_k$ (for $k=1,\ldots,E$),
but with different overall shifts $\Delta s$:
\[
s_{10,2}=(1,2,2,3,3,4,4,4,5,5,6,6,6,7,7,8,8,9)^{+\mathbf{4/8}}
\]
The example demonstrates that $\Delta s$ represents an independent characteristic for the graph
which cannot be deduced from the values $s_k$ alone.
In that sense, all the numbers $d_k$ describing the perimeter configuration
in terms of consecutive propagators must be provided (together with the initial flavour $f_1$).

\medskip

\begin{figure}\centering
\begin{mpostfig}
interim xu:=1cm;
paths[1]:=((3,2)--(0,2)--(0,0)--(3,0)--(3,0.9)--(2,0.9){dir 180}..{dir 0}(2,1.1)--(3,1.1)--cycle) scaled 1xu;
fill paths[1] withgreyscale 0.8;
draw ((0.5,0)--(0.5,2)) scaled 1xu pensize(0.1pt) withcolor 0.7[white,coltwo];
draw ((1.5,0)--(1.5,2)) scaled 1xu pensize(0.1pt) withcolor 0.7[white,coltwo];
draw ((2.5,0)--(2.5,0.9)) scaled 1xu pensize(0.1pt) withcolor 0.7[white,coltwo];
draw ((2.5,1.1)--(2.5,2)) scaled 1xu pensize(0.1pt) withcolor 0.7[white,coltwo];
draw ((0,1.5)--(3,1.5)) scaled 1xu pensize(0.1pt) withcolor 0.7[white,colone];
draw ((0,0.5)--(3,0.5)) scaled 1xu pensize(0.1pt) withcolor 0.7[white,colone];
draw paths[1] pensize(1pt);
midarrow(paths[1],1/3);
\end{mpostfig}
\caption{The fishnet shape corresponding to \protect\figref{fig:unique}.}
\label{fig:shape}
\end{figure}

Let us formulate two maximal uniqueness propositions
subject to what we have learned
from the above examples of non-unique graphs:
\begin{enumerate}
\item
A bi-scalar graph whose loops are all squares
is uniquely identified by its perimeter configuration $(s_j)$.
Such a graph is invariant under the conformal Yangian.
\item
A fishnet graph is uniquely identified by its flavour configuration $(f_j)$.
Such a graph is the single contribution to the corresponding bi-scalar correlator.
\end{enumerate}
In the following we will provide justifications for them.
To that end, we introduce the “fishnet shape” of a bi-scalar graph:
The fishnet shape is a polygonal curve drawn on a fishnet plane (square lattice).
For each external leg of flavour $(1,2,3,4)$,
the curve proceeds by a unit step in the (left, down, right, up) direction, respectively.
Charge conservation implies that the curve closes.
In \figref{fig:shape}, we display a sample fishnet shape
corresponding to the flavour configuration \eqref{eq:unique}.

\medskip

\begin{figure}\centering
\begin{mpostfig}
interim xu:=1cm;
paths[1]:=((2,0)--(2,1)--(2,2)--(2,3)--(3,3)--(3,2)--(2,2)--(1,2)--(0,2)--(0,1)
         --(1,1)--(2,1)--(3,1)--(4,1)--(4,2)--(4,3)--(4,4)--(3,4)--(2,4)--(1,4)
         --(1,3)--(1,2)--(1,1)--(1,0)--cycle) scaled 1xu;
paths[2]:=((2.3,2.5)--(2.3,0.5)) scaled 1xu;
fill subpath(5.7,11.3) of paths[1]--cycle withgreyscale 0.9;
fill subpath(11.3,29.7) of paths[1]--cycle withgreyscale 0.8;
draw ((1.5,0)--(1.5,4)) scaled 1xu pensize(0.1pt) withcolor 0.7[white,coltwo];
draw ((2.5,3)--(2.5,4)) scaled 1xu pensize(0.1pt) withcolor 0.7[white,coltwo];
draw ((2.5,1)--(2.5,2)) scaled 1xu pensize(0.1pt) withcolor 0.7[white,coltwo];
draw ((3.5,1)--(3.5,4)) scaled 1xu pensize(0.1pt) withcolor 0.7[white,coltwo];
draw ((1,3.5)--(4,3.5)) scaled 1xu pensize(0.1pt) withcolor 0.7[white,colone];
draw ((1,1.43)--(2,1.43)) scaled 1xu pensize(0.1pt) withcolor 0.7[white,colone];
draw ((1,0.5)--(2,0.5)) scaled 1xu pensize(0.1pt) withcolor 0.7[white,colone];
draw ((1,2.5)--(2,2.5)) scaled 1xu pensize(0.1pt) withcolor 0.7[white,colone];
draw ((3,2.5)--(4,2.5)) scaled 1xu pensize(0.1pt) withcolor 0.7[white,colone];
draw ((0,1.5)--(2.3,1.5)) scaled 1xu pensize(0.1pt) dashed withdots scaled 0.3 withcolor 0.7[white,colone];
draw ((2.3,1.5)--(4,1.5)) scaled 1xu pensize(0.1pt) withcolor 0.7[white,colone];
draw ((1.57,1)--(1.57,2)) scaled 1xu pensize(0.1pt) dashed withdots scaled 0.3 withcolor 0.7[white,coltwo];
draw ((0.5,1)--(0.5,2)) scaled 1xu pensize(0.1pt) dashed withdots scaled 0.3 withcolor 0.7[white,coltwo];
draw subpath(5.7,11.3) of paths[1] pensize(1pt) dashed withdots scaled 0.5;
draw subpath(11.3,29.7) of paths[1] pensize(1pt);
midarrow(paths[1],1.5);
draw paths[2] pensize(1pt) dashed evenly withcolor 0.8red;
drawdot(point 0 of paths[2],3pt) withcolor 0.8red;
drawdot(point 1 of paths[2],3pt) withcolor 0.8red;
\end{mpostfig}
\caption{A globally overlapping fishnet shape
with branch cut resolution.}
\label{fig:overlap}
\end{figure}

\begin{figure}\centering
\begin{mpostfig}[opt=align]
interim xu:=1.5cm;
paths[1]:=((0,-0.5)--(0,-0.1){dir 90}..{dir 180}(-0.1,0)--(-0.5,0)) scaled 1xu;
fill unitsquare shifted (-0.5,-0.5) scaled 1xu withgreyscale 1;
fill paths[1]--(-0.5xu,-0.5xu)--cycle withgreyscale 0.8; 
draw ((-0.25,-0.5)--(-0.25,0)) scaled 1xu pensize(0.1pt) withcolor 0.7[white,coltwo];
draw ((-0.5,-0.25)--(0,-0.25)) scaled 1xu pensize(0.1pt) withcolor 0.7[white,colone];
draw paths[1] pensize(1pt);
midarrow(paths[1],0.5);
\end{mpostfig}
\qquad
\begin{mpostfig}[opt=align]
interim xu:=1.5cm;
paths[1]:=((0,-0.5)--(0,0.5)) scaled 1xu;
fill unitsquare shifted (-0.5,-0.5) scaled 1xu withgreyscale 1;
fill paths[1]--(-0.5xu,0.5xu)--(-0.5xu,-0.5xu)--cycle withgreyscale 0.8; 
draw ((-0.25,-0.5)--(-0.25,+0.5)) scaled 1xu pensize(0.1pt) withcolor 0.7[white,coltwo];
draw ((-0.5,-0.25)--(0,-0.25)) scaled 1xu pensize(0.1pt) withcolor 0.7[white,colone];
draw ((-0.5,+0.25)--(0,+0.25)) scaled 1xu pensize(0.1pt) withcolor 0.7[white,colone];
draw paths[1] pensize(1pt);
midarrow(paths[1],0.5);
\end{mpostfig}
\qquad
\begin{mpostfig}[opt=align]
interim xu:=1.5cm;
paths[1]:=((0,-0.5)--(0,-0.1){dir 90}..{dir 0}(0.1,0)--(0.5,0)) scaled 1xu;
fill unitsquare shifted (-0.5,-0.5) scaled 1xu withgreyscale 1;
fill paths[1]--(0.5xu,0.5xu)--(-0.5xu,0.5xu)--(-0.5xu,-0.5xu)--cycle withgreyscale 0.8; 
draw ((-0.25,-0.5)--(-0.25,+0.5)) scaled 1xu pensize(0.1pt) withcolor 0.7[white,coltwo];
draw ((+0.25,+0.5)--(+0.25,0)) scaled 1xu pensize(0.1pt) withcolor 0.7[white,coltwo];
draw ((-0.5,-0.25)--(0,-0.25)) scaled 1xu pensize(0.1pt) withcolor 0.7[white,colone];
draw ((-0.5,+0.25)--(+0.5,+0.25)) scaled 1xu pensize(0.1pt) withcolor 0.7[white,colone];
draw paths[1] pensize(1pt);
midarrow(paths[1],0.5);
\end{mpostfig}
\qquad
\begin{mpostfig}[opt=align]
interim xu:=1.5cm;
paths[1]:=((-0.05,-0.5)--(-0.05,0){dir 90}..{dir -90}(0.05,0)--(0.05,-0.5)) scaled 1xu;
fill unitsquare shifted (-0.5,-0.5) scaled 1xu withgreyscale 1;
fill paths[1]--(0.5xu,-0.5xu)--(0.5xu,0.5xu)--(-0.5xu,0.5xu)--(-0.5xu,-0.5xu)--cycle withgreyscale 0.8; 
draw ((-0.25,-0.5)--(-0.25,+0.5)) scaled 1xu pensize(0.1pt) withcolor 0.7[white,coltwo];
draw ((+0.25,+0.5)--(+0.25,-0.5)) scaled 1xu pensize(0.1pt) withcolor 0.7[white,coltwo];
draw ((-0.5,-0.25)--(-0.05,-0.25)) scaled 1xu pensize(0.1pt) withcolor 0.7[white,colone];
draw ((0.05,-0.25)--(0.5,-0.25)) scaled 1xu pensize(0.1pt) withcolor 0.7[white,colone];
draw ((-0.5,+0.25)--(+0.5,+0.25)) scaled 1xu pensize(0.1pt) withcolor 0.7[white,colone];
draw paths[1] pensize(1pt);
midarrow(paths[1],0.5);
\end{mpostfig}
\qquad
\begin{mpostfig}[opt=align]
interim xu:=1.5cm;
paths[1]:=((0,-0.5)--(0,0.1){dir 90}..(0.1,0.2){dir 0}..(0.2,0.1){dir -90}..{dir 180}(0.1,0)--(-0.5,0)) scaled 1xu;
paths[2]:=((0.1,0.1)--(0.1,-0.5)) scaled xu;
fill unitsquare shifted (-0.5,-0.5) scaled 1xu withgreyscale 1;
fill (subpath(4,5) of paths[1])--(-0.5xu,-0.5xu)--(0.1xu,-0.5xu)--cycle withgreyscale 0.9; 
fill (subpath(0,4) of paths[1])--(0.1xu,-0.5xu)--(0.5xu,-0.5xu)--(0.5xu,0.5xu)--(-0.5xu,0.5xu)--(-0.5xu,-0.5xu)--cycle withgreyscale 0.8; 
draw ((-0.25,-0.5)--(-0.25,+0.5)) scaled 1xu pensize(0.1pt) withcolor 0.7[white,coltwo];
draw ((+0.25,+0.5)--(+0.25,-0.5)) scaled 1xu pensize(0.1pt) withcolor 0.7[white,coltwo];
draw ((-0.5,-0.3)--(-0,-0.3)) scaled 1xu pensize(0.1pt) withcolor 0.7[white,colone];
draw ((0.1,-0.25)--(0.5,-0.25)) scaled 1xu pensize(0.1pt) withcolor 0.7[white,colone];
draw ((-0.2,-0.5)--(-0.2,0)) scaled 1xu pensize(0.1pt) dashed withdots scaled 0.3 withcolor 0.7[white,coltwo];
draw ((-0.5,-0.25)--(0.1,-0.25)) scaled 1xu pensize(0.1pt) dashed withdots scaled 0.3 withcolor 0.7[white,colone];
draw ((-0.5,+0.25)--(+0.5,+0.25)) scaled 1xu pensize(0.1pt) withcolor 0.7[white,colone];
draw paths[2] pensize (1pt) dashed evenly withcolor 0.8red;
drawdot(point 0 of paths[2],3pt) withcolor 0.8red;
draw subpath(0,4) of paths[1] pensize(1pt);
draw subpath(4,5) of paths[1] pensize(1pt) dashed withdots scaled 0.5;
midarrow(paths[1],0.5);
\end{mpostfig}
\qquad
\begin{mpostfig}[opt=align]
interim xu:=1.5cm;
paths[1]:=((0,-0.5)--(0,0){dir 90}..(0.1,0.1){dir 0}..(0.2,0){dir -90}..{dir 180}(0.1,-0.1)..{dir 90}(0,0)--(0,0.5)) scaled 1xu;
paths[2]:=((0.1,0)--(0.1,-0.5)) scaled xu;
fill unitsquare shifted (-0.5,-0.5) scaled 1xu withgreyscale 1;
fill (subpath(4,5) of paths[1])--(-0.5xu,0.5xu)--(-0.5xu,-0.5xu)--(0.1xu,-0.5xu)--cycle withgreyscale 0.9; 
fill (subpath(0,4) of paths[1])--(0.1xu,-0.5xu)--(0.5xu,-0.5xu)--(0.5xu,0.5xu)--(-0.5xu,0.5xu)--(-0.5xu,-0.5xu)--cycle withgreyscale 0.8; 
draw ((-0.25,-0.5)--(-0.25,+0.5)) scaled 1xu pensize(0.1pt) withcolor 0.7[white,coltwo];
draw ((+0.25,+0.5)--(+0.25,-0.5)) scaled 1xu pensize(0.1pt) withcolor 0.7[white,coltwo];
draw ((-0.5,-0.3)--(-0,-0.3)) scaled 1xu pensize(0.1pt) withcolor 0.7[white,colone];
draw ((0.1,-0.25)--(0.5,-0.25)) scaled 1xu pensize(0.1pt) withcolor 0.7[white,colone];
draw ((-0.2,-0.5)--(-0.2,+0.5)) scaled 1xu pensize(0.1pt) dashed withdots scaled 0.3 withcolor 0.7[white,coltwo];
draw ((-0.5,-0.25)--(0.1,-0.25)) scaled 1xu pensize(0.1pt) dashed withdots scaled 0.3 withcolor 0.7[white,colone];
draw ((-0.5,+0.25)--(0.0,+0.25)) scaled 1xu pensize(0.1pt) dashed withdots scaled 0.3 withcolor 0.7[white,colone];
draw ((-0.5,+0.2)--(+0.5,+0.2)) scaled 1xu pensize(0.1pt) withcolor 0.7[white,colone];
draw paths[2] pensize (1pt) dashed evenly withcolor 0.8red;
drawdot(point 0 of paths[2],3pt) withcolor 0.8red;
draw subpath(0,4) of paths[1] pensize(1pt);
draw subpath(4,6) of paths[1] pensize(1pt) dashed withdots scaled 0.5;
midarrow(paths[1],0.5);
\end{mpostfig}
\qquad
\ldots
\caption{Turns of the fishnet shape by $+90^\circ, 0^\circ, -90^\circ, -180^\circ, -270^\circ, -360^\circ, \ldots$
corresponding to $d_j=0,1,2,3,4,5,\ldots$ adjacent perimeter propagators.}
\label{fig:turns}
\end{figure}

First, we consider a bi-scalar graph whose loops are all squares, $N_j=4$.
The perimeter configuration identifies an evaluation parameter sequence $s_j$
such that $\Delta s=4$,
and it prescribes a fishnet shape which can be self-intersecting.
In order to remove the self-intersections
we draw the fishnet shape on a fishnet plane with suitably arranged sheets.
The branching and inter-connectivity of the sheets is determined 
by the global overlapping of parts of the fishnet shape
as well as by the local perimeter configuration:
If two planar parts of the fishnet shape overlap,
they need to be separated by introducing a branch cut between them
such that the parts bypass each other on different sheets,
see \figref{fig:overlap}.
The perimeter configuration may require further branchings
due to turns about some angles $(1-d_j){\cdot}90^\circ$ exceeding $-180^\circ$.
In order to resolve this situation, a branch cut must extend right to the corner 
of the fishnet shape such that the curve can locally wind back onto itself,
see \figref{fig:turns}.
Importantly, the condition $\Delta s=4$ implies that the fishnet curve takes just one full overall turn. 
This implies that all branch points reside on the exterior of the curve, 
there is no net winding of the curve around them,
and for each crossing of a branch cut there has to be a corresponding opposite crossing.
The bi-scalar graph can be reconstructed from this information:
We cut the fishnet shape into several planar pieces at the branch cuts
while remembering the interfaces between the pieces across the branch cuts.
Each piece cuts a bi-scalar graph out of a fishnet plane.
These pieces are patched according to the information
from cutting the fishnet shape at the branch cuts.
We thus obtain a fishnet graph defined uniquely by the procedure.
Another consequence is that the fishnet shape is contractible
to a point on the properly branched fishnet.
Contractibility is an essential prerequisite
for the inductive procedure described in \cite{Chicherin:2017cns}
to show Yangian invariance of a bi-scalar graph.
This procedure consists of a sequence of moves
which chip off one unit square area
from the interior of the fishnet shape at a time.
It can be applied equally to the layered bi-scalar graph
thus ensuring its Yangian invariance.

As an aside, we remark that 
the above considerations imply a simple relationship for the evaluation parameters
$s_{j,k}$ for the two ends $j,k$ of an uninterrupted flavour line:
\[
s_k=s_j+2
\& 
\text{ for }1\leq j<k\leq E
\]
The evaluation parameter increases by precisely two units 
between two legs connected by an internal line in the forward direction
around the perimeter of the graph.
Interestingly, the relationship can work consistently with cyclicity
in the opposite direction
\[
s_k=s_{j-E}+2=s_j-\Delta s+2=s_j-2
\&
\text{ for }1\leq k<j\leq E
\],
if and only if the quasi-periodicity of evaluation parameters is $\Delta s=4$.
All sample graphs with $\Delta s=4$ indeed satisfy the relation
in the forward and backward directions alike.
Importantly, the move described in \figref{fig:octagonflip}
either changes some of the evaluation parameters of the involved legs
or it implies $\Delta s=8$.
Hence, it has no impact on the special case $\Delta s=4$.

\medskip

We further note that the case $\Delta s>4$ implies that the fishnet shape 
has a net winding of $\Delta s{\cdot}90^\circ>360^\circ$. 
This winding can only be caused by branch points 
with a surplus angle residing in the interior of the fishnet shape.
These imply that the fishnet shape traverses the corresponding branch cuts 
several times in the same direction before returning to the original sheet.
Altogether, such a fishnet shape cannot be contracted to a point on the associated fishnet with sheets,
and the above reconstruction of the corresponding graph does not apply.
Indeed, we have already encountered examples of non-unique bi-scalar graphs when $\Delta s>4$.
Furthermore, the internal surplus angles indicate loops with more than 4 sides;
these spoil the invariance of the bi-scalar graph under the conformal Yangian algebra.

\medskip

Let us finally address the uniqueness for fishnet graphs
which can be cut out from a single large piece of fishnet by cutting the links
along a closed curve without self-intersections.
The fishnet shape for such a graph describes where the links on the larger fishnet are cut.
For that reason it is a planar curve without self-intersections.
\unskip\footnote{Some parts of the curve may run in parallel,
see the case $180^\circ$ in \figref{fig:turns}.
Here, one should assume that there remains a small strip of intermediate area
which is not enclosed by the curve.}
Consequently, the numbers $d_j$ can only take the values $0,1,2,3$,
and these are linked uniquely to the flavour sequence $f_j$.
Hence, the flavour sequence $f_j$ describes a unique fishnet shape.
Cutting a large fishnet along the fishnet shape reproduces the fishnet graph.
The relevant question is whether any other sequence of numbers $d_j$
compatible with the flavours $f_j$ might describe a different bi-scalar graph:
However, such a change cannot lead to an actual graph
because the $d_j$ may only be increased by positive integer multiples of $4$,
and this would necessarily lead to a decrease of $\Delta s$ according to \eqref{eq:svsd}
which is already at its minimum value $4$ for a fishnet graph.
Fishnet graphs indeed provide unique contributions to the corresponding bi-scalar correlators.

\providecommand{\nbbsturl}[1]{\texttt{\def\tild{\char126}\let~\tild#1}}

\begin{bibtex}[\jobname]

@article{Beisert:2017pnr,
    author = "Beisert, Niklas and Garus, Aleksander and Rosso, Matteo",
    title = "{Yangian Symmetry and Integrability of Planar N=4 Supersymmetric Yang-Mills Theory}",
    eprint = "1701.09162",
    archivePrefix = "arXiv",
    primaryClass = "hep-th",
    reportNumber = "HU-EP-17-03",
    doi = "10.1103/PhysRevLett.118.141603",
    journal = "Phys. Rev. Lett.",
    volume = "118",
    number = "14",
    pages = "141603",
    year = "2017"
}

@article{Beisert:2018zxs,
    author = "Beisert, Niklas and Garus, Aleksander and Rosso, Matteo",
    title = "{Yangian Symmetry for the Action of Planar N=4 Super Yang-Mills and N=6 Super Chern-Simons Theories}",
    eprint = "1803.06310",
    archivePrefix = "arXiv",
    primaryClass = "hep-th",
    reportNumber = "HU-EP-18/08, HU-EP-18-08",
    doi = "10.1103/PhysRevD.98.046006",
    journal = "Phys. Rev. D",
    volume = "98",
    number = "4",
    pages = "046006",
    year = "2018"
}

@article{Beisert:2018ijg,
    author = "Beisert, Niklas and Garus, Aleksander",
    title = "{Yangian Algebra and Correlation Functions in Planar Gauge Theories}",
    eprint = "1804.09110",
    archivePrefix = "arXiv",
    primaryClass = "hep-th",
    doi = "10.21468/SciPostPhys.5.2.018",
    journal = "SciPost Phys.",
    volume = "5",
    number = "2",
    pages = "018",
    year = "2018"
}

@article{Garus:2017bgl,
    author = "Garus, Aleksander",
    title = "{Untwisting the symmetries of $\beta$-deformed Super-Yang--Mills}",
    eprint = "1707.04128",
    archivePrefix = "arXiv",
    primaryClass = "hep-th",
    doi = "10.1007/JHEP10(2017)007",
    journal = "JHEP",
    volume = "10",
    pages = "007",
    year = "2017"
}

@article{Beisert:2024wqq,
    author = {Beisert, Niklas and K{\"o}nig, Benedikt},
    title = "{Yangian form-alism for planar gauge theories}",
    eprint = "2411.16176",
    archivePrefix = "arXiv",
    primaryClass = "hep-th",
    doi = "10.1063/5.0253127",
    journal = "J. Math. Phys.",
    volume = "66",
    number = "6",
    pages = "062301",
    year = "2025"
}

@article{Drummond:2009fd,
    author = "Drummond, James M. and Henn, Johannes M. and Plefka, Jan",
    editor = "Liu, Feng and Xiao, Zhigang and Zhuang, Pengfei",
    title = "{Yangian symmetry of scattering amplitudes in N=4 super Yang-Mills theory}",
    eprint = "0902.2987",
    archivePrefix = "arXiv",
    primaryClass = "hep-th",
    reportNumber = "HU-EP-09-06, LAPTH-1308-09",
    doi = "10.1088/1126-6708/2009/05/046",
    journal = "JHEP",
    volume = "05",
    pages = "046",
    year = "2009"
}

@article{ReshetikhinTwist,
     author    = "Reshetikhin, N. {\relax Yu}.",
     title     = "Multiparametric quantum groups and twisted quasitriangular
                  Hopf algebras",
     journal   = "Lett. Math. Phys.",
     volume    = "20",
     year      = "1990",
     pages     = "331",
     doi = "10.1007/BF00626530"
}

@article{Leigh:1995ep,
    author = "Leigh, Robert G. and Strassler, Matthew J.",
    title = "{Exactly marginal operators and duality in four-dimensional N=1 supersymmetric gauge theory}",
    eprint = "hep-th/9503121",
    archivePrefix = "arXiv",
    reportNumber = "RU-95-2",
    doi = "10.1016/0550-3213(95)00261-P",
    journal = "Nucl. Phys. B",
    volume = "447",
    pages = "95--136",
    year = "1995"
}

@article{Drinfeld:1989st,
    author = "Drinfeld, V. G.",
    title = "Quasi Hopf algebras",
    journal = "Alg. Anal.",
    volume = "1",
    pages = "114--148",
    year = "1989",
    url = {https://www.mathnet.ru/eng/aa/v1/i6/p114}
}

@article{Lunin:2005jy,
    author = "Lunin, Oleg and Maldacena, Juan Martin",
    title = "{Deforming field theories with U(1) $\times$ U(1) global symmetry and their gravity duals}",
    eprint = "hep-th/0502086",
    archivePrefix = "arXiv",
    doi = "10.1088/1126-6708/2005/05/033",
    journal = "JHEP",
    volume = "05",
    pages = "033",
    year = "2005"
}

@article{Frolov:2005dj,
    author = "Frolov, Sergey",
    title = "{Lax pair for strings in Lunin-Maldacena background}",
    eprint = "hep-th/0503201",
    archivePrefix = "arXiv",
    doi = "10.1088/1126-6708/2005/05/069",
    journal = "JHEP",
    volume = "05",
    pages = "069",
    year = "2005"
}

@article{Beisert:2005if,
    author = "Beisert, N. and Roiban, R.",
    title = "{Beauty and the twist: The Bethe ansatz for twisted N=4 SYM}",
    eprint = "hep-th/0505187",
    archivePrefix = "arXiv",
    reportNumber = "PUTP-2162",
    doi = "10.1088/1126-6708/2005/08/039",
    journal = "JHEP",
    volume = "08",
    pages = "039",
    year = "2005"
}

@article{Gurdogan:2015csr,
    author = {Gürdoğan, {\relax Ö}mer and Kazakov, Vladimir},
    title = "{New Integrable 4D Quantum Field Theories from Strongly Deformed Planar N=4 Supersymmetric Yang-Mills Theory}",
    eprint = "1512.06704",
    archivePrefix = "arXiv",
    primaryClass = "hep-th",
    doi = "10.1103/PhysRevLett.117.201602",
    journal = "Phys. Rev. Lett.",
    volume = "117",
    number = "20",
    pages = "201602",
    year = "2016"
}

@article{Sieg:2016vap,
    author = "Sieg, Christoph and Wilhelm, Matthias",
    title = "{On a CFT limit of planar $\gamma_i$-deformed N=4 SYM theory}",
    eprint = "1602.05817",
    archivePrefix = "arXiv",
    primaryClass = "hep-th",
    reportNumber = "HU-MATHEMATIK-2016-02, {\textbackslash}-HU-EP-16-06, HU-EP-16-06",
    doi = "10.1016/j.physletb.2016.03.004",
    journal = "Phys. Lett. B",
    volume = "756",
    pages = "118--120",
    year = "2016"
}

@article{Caetano:2016ydc,
    author = {Caetano, João and Gürdoğan, {\relax Ö}mer and Kazakov, Vladimir},
    title = "{Chiral limit of N=4 SYM and ABJM and integrable Feynman graphs}",
    eprint = "1612.05895",
    archivePrefix = "arXiv",
    primaryClass = "hep-th",
    doi = "10.1007/JHEP03(2018)077",
    journal = "JHEP",
    volume = "03",
    pages = "077",
    year = "2018"
}

@article{Chicherin:2017cns,
    author = {Chicherin, Dmitry and Kazakov, Vladimir and Loebbert, Florian and Müller, Dennis and Zhong, De-liang},
    title = "{Yangian Symmetry for Bi-Scalar Loop Amplitudes}",
    eprint = "1704.01967",
    archivePrefix = "arXiv",
    primaryClass = "hep-th",
    reportNumber = "HU-EP-17-09, MITP-17-022, LPTENS-17-07",
    doi = "10.1007/JHEP05(2018)003",
    journal = "JHEP",
    volume = "05",
    pages = "003",
    year = "2018"
}

@article{Beisert:2010jr,
    author = "Beisert, Niklas and others",
    title = "{Review of AdS/CFT Integrability: An Overview}",
    eprint = "1012.3982",
    archivePrefix = "arXiv",
    primaryClass = "hep-th",
    reportNumber = "AEI-2010-175, CERN-PH-TH-2010-306, HU-EP-10-87, HU-MATH-2010-22, KCL-MTH-10-10, UMTG-270, UUITP-41-10",
    doi = "10.1007/s11005-011-0529-2",
    journal = "Lett. Math. Phys.",
    volume = "99",
    pages = "3--32",
    year = "2012"
}

@article{Drinfel'd:1985,
    author = "Drinfel'd, Vladimir Gershonovich",
    title = "Hopf algebras and the quantum Yang–Baxter equation",
    journal   = "Sov. Math. Dokl.",
    volume    = "32",
    pages     = "254-258",
    year      = "1985"
}

@article{Drinfel'd:1988,
    author = "Drinfel'd, Vladimir Gershonovich",
    title = "Quantum groups",
    doi = "10.1007/BF01247086",
    journal = "J. Sov. Math.",
    volume = "41",
    issue = "2",
    pages = "898--915",
    year = "1988"
}

@article{Dolan:2004ps,
	author    = "Dolan, Louise and Nappi, Chiara R. and Witten, Edward",
	title     = "Yangian symmetry in D=4 superconformal Yang--Mills theory",
	booktitle = "Quantum Theory and Symmetries",
	series    = "Proceedings of the 3rd International Symposium, Cincinnati, USA, 10-14 September 2003",
	editor    = "Argyres, P. C. and others",
	publisher = "World Scientific",
	year      = "2004",
	address   = "Singapore",
    doi = "10.1142/9789812702340_0036",
    pages = "300--315",
	eprint    = "hep-th/0401243",
	archivePrefix = "arXiv",
}

@article{Dolan:2003uh,
    author = "Dolan, Louise and Nappi, Chiara R. and Witten, Edward",
    title = "{A Relation between approaches to integrability in superconformal Yang-Mills theory}",
    eprint = "hep-th/0308089",
    archivePrefix = "arXiv",
    doi = "10.1088/1126-6708/2003/10/017",
    journal = "JHEP",
    volume = "10",
    pages = "017",
    year = "2003"
}

@article{Chicherin:2017frs,
    author = {Chicherin, Dmitry and Kazakov, Vladimir and Loebbert, Florian and Müller, Dennis and Zhong, De-liang},
    title = "{Yangian Symmetry for Fishnet Feynman Graphs}",
    eprint = "1708.00007",
    archivePrefix = "arXiv",
    primaryClass = "hep-th",
    reportNumber = "MITP-17-049, HU-EP-17-20, LPTENS-17-32",
    doi = "10.1103/PhysRevD.96.121901",
    journal = "Phys. Rev. D",
    volume = "96",
    number = "12",
    pages = "121901",
    year = "2017"
}

@article{Zamolodchikov:1980mb,
    author = "Zamolodchikov, A. B.",
    title = "{`Fishing-net' diagrams as a completely integrable system}",
    doi = "10.1016/0370-2693(80)90547-X",
    journal = "Phys. Lett. B",
    volume = "97",
    pages = "63--66",
    year = "1980"
}

@article{Loebbert:2025abz,
    author = {Loebbert, Florian and Rüenaufer, Lucas and Stawinski, Sven F.},
    title = "{Nonlocal Symmetries of Planar Feynman Integrals}",
    eprint = "2505.05550",
    archivePrefix = "arXiv",
    primaryClass = "hep-th",
    reportNumber = "BONN-TH-2025-20",
    doi = "10.1103/fvzk-dp1b",
    journal = "Phys. Rev. Lett.",
    volume = "135",
    number = "15",
    pages = "151603",
    year = "2025"
}

@article{Frolov:2005ty,
    author = "Frolov, S. A. and Roiban, R. and Tseytlin, Arkady A.",
    title = "{Gauge-string duality for superconformal deformations of N=4 super Yang-Mills theory}",
    eprint = "hep-th/0503192",
    archivePrefix = "arXiv",
    doi = "10.1088/1126-6708/2005/07/045",
    journal = "JHEP",
    volume = "07",
    pages = "045",
    year = "2005"
}

@article{Frolov:2005iq,
    author = "Frolov, S. A. and Roiban, R. and Tseytlin, Arkady A.",
    title = "{Gauge-string duality for (non)supersymmetric deformations of N=4 super Yang-Mills theory}",
    eprint = "hep-th/0507021",
    archivePrefix = "arXiv",
    reportNumber = "PUPT-2167, AEI-2005-118",
    doi = "10.1016/j.nuclphysb.2005.10.004",
    journal = "Nucl. Phys. B",
    volume = "731",
    pages = "1--44",
    year = "2005"
}

@article{Chicherin:2022nqq,
    author = "Chicherin, Dmitry and Korchemsky, Gregory P.",
    title = "{The SAGEX review on scattering amplitudes Chapter 9: Integrability of amplitudes in fishnet theories}",
    eprint = "2203.13020",
    archivePrefix = "arXiv",
    primaryClass = "hep-th",
    reportNumber = "SAGEX-22-10",
    doi = "10.1088/1751-8121/ac8c72",
    journal = "J. Phys. A",
    volume = "55",
    number = "44",
    pages = "443010",
    year = "2022"
}

@article{Loebbert:2020tje,
    author = "Loebbert, Florian and Miczajka, Julian",
    title = "{Massive Fishnets}",
    eprint = "2008.11739",
    archivePrefix = "arXiv",
    primaryClass = "hep-th",
    reportNumber = "HU-EP-20/21",
    doi = "10.1007/JHEP12(2020)197",
    journal = "JHEP",
    volume = "12",
    pages = "197",
    year = "2020"
}

@article{Kazakov:2023nyu,
    author = "Kazakov, Vladimir and Levkovich-Maslyuk, Fedor and Mishnyakov, Victor",
    title = "{Integrable Feynman graphs and Yangian symmetry on the loom}",
    eprint = "2304.04654",
    archivePrefix = "arXiv",
    primaryClass = "hep-th",
    doi = "10.1007/JHEP06(2025)104",
    journal = "JHEP",
    volume = "06",
    pages = "104",
    year = "2025"
}

@article{Kazakov:2022dbd,
    author = "Kazakov, Vladimir and Olivucci, Enrico",
    title = "{The loom for general fishnet CFTs}",
    eprint = "2212.09732",
    archivePrefix = "arXiv",
    primaryClass = "hep-th",
    doi = "10.1007/JHEP06(2023)041",
    journal = "JHEP",
    volume = "06",
    pages = "041",
    year = "2023"
}

@article{Pittelli:2019ceq,
    author = "Pittelli, Antonio and Preti, Michelangelo",
    title = "{Integrable fishnet from $\gamma$-deformed N=2 quivers}",
    eprint = "1906.03680",
    archivePrefix = "arXiv",
    primaryClass = "hep-th",
    reportNumber = "NORDITA 2019-056, UUITP -- 21/19, UUITP-21/19",
    doi = "10.1016/j.physletb.2019.134971",
    journal = "Phys. Lett. B",
    volume = "798",
    pages = "134971",
    year = "2019"
}

@article{Basso:2015zoa,
    author = "Basso, Benjamin and Komatsu, Shota and Vieira, Pedro",
    title = "{Structure Constants and Integrable Bootstrap in Planar N=4 SYM Theory}",
    eprint = "1505.06745",
    archivePrefix = "arXiv",
    primaryClass = "hep-th",
    month = "5",
    year = "2015"
}

@article{Gromov:2017cja,
    author = "Gromov, Nikolay and Kazakov, Vladimir and Korchemsky, Gregory and Negro, Stefano and Sizov, Grigory",
    title = "{Integrability of Conformal Fishnet Theory}",
    eprint = "1706.04167",
    archivePrefix = "arXiv",
    primaryClass = "hep-th",
    doi = "10.1007/JHEP01(2018)095",
    journal = "JHEP",
    volume = "01",
    pages = "095",
    year = "2018"
}

@article{Kazakov:2018ugh,
    author = "Kazakov, Vladimir",
    editor = "Ge, Mo-Lin and Niemi, Antti J. and Phua, Kok Khoo and Takhtajan, Leon A.",
    title = "{Quantum Spectral Curve of $\gamma$-twisted N=4 SYM theory and fishnet CFT}",
    eprint = "1802.02160",
    archivePrefix = "arXiv",
    primaryClass = "hep-th",
    reportNumber = "LPTENS-18-02, LPTENS-18/02",
    doi = "10.1142/9789813233867_0016",
    pages = "293--342",
    year = "2018"
}

@article{Ipsen:2018fmu,
    author = "Ipsen, Asger C. and Staudacher, Matthias and Zippelius, Leonard",
    title = "{The one-loop spectral problem of strongly twisted N=4 Super Yang-Mills theory}",
    eprint = "1812.08794",
    archivePrefix = "arXiv",
    primaryClass = "hep-th",
    reportNumber = "HU-Mathematik-2018-11, HU-EP-18/39",
    doi = "10.1007/JHEP04(2019)044",
    journal = "JHEP",
    volume = "04",
    pages = "044",
    year = "2019"
}

@article{Loebbert:2020glj,
    author = {Loebbert, Florian and Miczajka, Julian and Müller, Dennis and Münkler, Hagen},
    title = "{Yangian Bootstrap for Massive Feynman Integrals}",
    eprint = "2010.08552",
    archivePrefix = "arXiv",
    primaryClass = "hep-th",
    reportNumber = "HU-EP-20/27",
    doi = "10.21468/SciPostPhys.11.1.010",
    journal = "SciPost Phys.",
    volume = "11",
    pages = "010",
    year = "2021"
}

@article{Loebbert:2019vcj,
    author = {Loebbert, Florian and Müller, Dennis and Münkler, Hagen},
    title = "{Yangian Bootstrap for Conformal Feynman Integrals}",
    eprint = "1912.05561",
    archivePrefix = "arXiv",
    primaryClass = "hep-th",
    reportNumber = "HU-EP-19/39",
    doi = "10.1103/PhysRevD.101.066006",
    journal = "Phys. Rev. D",
    volume = "101",
    number = "6",
    pages = "066006",
    year = "2020"
}

@article{Corcoran:2020epz,
    author = "Corcoran, Luke and Loebbert, Florian and Miczajka, Julian and Staudacher, Matthias",
    title = "{Minkowski Box from Yangian Bootstrap}",
    eprint = "2012.07852",
    archivePrefix = "arXiv",
    primaryClass = "hep-th",
    reportNumber = "HU-Mathematik-2020-07, HU-EP-20/38, SAGEX-20-27-E",
    doi = "10.1007/JHEP04(2021)160",
    journal = "JHEP",
    volume = "04",
    pages = "160",
    year = "2021"
}

@article{Olivucci:2021cfy,
    author = "Olivucci, Enrico",
    title = "{Hexagonalization of Fishnet integrals. Part I. Mirror excitations}",
    eprint = "2107.13035",
    archivePrefix = "arXiv",
    primaryClass = "hep-th",
    doi = "10.1007/JHEP11(2021)204",
    journal = "JHEP",
    volume = "11",
    pages = "204",
    year = "2021"
}

@article{Basso:2018cvy,
    author = "Basso, Benjamin and Caetano, João and Fleury, Thiago",
    title = "{Hexagons and Correlators in the Fishnet Theory}",
    eprint = "1812.09794",
    archivePrefix = "arXiv",
    primaryClass = "hep-th",
    doi = "10.1007/JHEP11(2019)172",
    journal = "JHEP",
    volume = "11",
    pages = "172",
    year = "2019"
}

@article{Corcoran:2021gda,
    author = "Corcoran, Luke and Loebbert, Florian and Miczajka, Julian",
    title = "{Yangian Ward identities for fishnet four-point integrals}",
    eprint = "2112.06928",
    archivePrefix = "arXiv",
    primaryClass = "hep-th",
    reportNumber = "HU-EP-21/54, SAGEX-21-38-E",
    doi = "10.1007/JHEP04(2022)131",
    journal = "JHEP",
    volume = "04",
    pages = "131",
    year = "2022"
}

@article{Faddeev:1994zg,
    author = "Faddeev, L. D. and Korchemsky, G. P.",
    title = "{High-energy QCD as a completely integrable model}",
    eprint = "hep-th/9404173",
    archivePrefix = "arXiv",
    reportNumber = "ITP-SB-94-14",
    doi = "10.1016/0370-2693(94)01363-H",
    journal = "Phys. Lett. B",
    volume = "342",
    pages = "311--322",
    year = "1995"
}

@article{Bombardelli:2016rwb,
    author = "Bombardelli, Diego and Cagnazzo, Alessandra and Frassek, Rouven and Levkovich-Maslyuk, Fedor and Loebbert, Florian and Negro, Stefano and Sz{\'e}cs{\'e}nyi, Istvan M. and Sfondrini, Alessandro and van Tongeren, Stijn J. and Torrielli, Alessandro",
    title = "{An integrability primer for the gauge-gravity correspondence: An introduction}",
    eprint = "1606.02945",
    archivePrefix = "arXiv",
    primaryClass = "hep-th",
    reportNumber = "CNRS-16-03, DCPT-16-19, DESY-16-083, DMUS-MP-16-09, HU-EP-16-13, NORDITA-2016-33, HU-MATH-16-08",
    doi = "10.1088/1751-8113/49/32/320301",
    journal = "J. Phys. A",
    volume = "49",
    number = "32",
    pages = "320301",
    year = "2016"
}

@article{Grabner:2017pgm,
    author = "Grabner, David and Gromov, Nikolay and Kazakov, Vladimir and Korchemsky, Gregory",
    title = "{Strongly $\gamma$-Deformed N=4 Supersymmetric Yang-Mills Theory as an Integrable Conformal Field Theory}",
    eprint = "1711.04786",
    archivePrefix = "arXiv",
    primaryClass = "hep-th",
    reportNumber = "KCL-MTH-17-04, LPTENS-17-31, IPHT-T17-171, LPTENS--17-31, IPHT--T17-171",
    doi = "10.1103/PhysRevLett.120.111601",
    journal = "Phys. Rev. Lett.",
    volume = "120",
    number = "11",
    pages = "111601",
    year = "2018"
}

@article{Gromov:2019jfh,
    author = "Gromov, Nikolay and Sever, Amit",
    title = "{The holographic dual of strongly $\gamma$-deformed N=4 SYM theory: derivation, generalization, integrability and discrete reparametrization symmetry}",
    eprint = "1908.10379",
    archivePrefix = "arXiv",
    primaryClass = "hep-th",
    reportNumber = "CERN-TH-2019-144",
    doi = "10.1007/JHEP02(2020)035",
    journal = "JHEP",
    volume = "02",
    pages = "035",
    year = "2020"
}

@article{Gromov:2019bsj,
    author = "Gromov, Nikolay and Sever, Amit",
    title = "{Quantum fishchain in AdS$_{5}$}",
    eprint = "1907.01001",
    archivePrefix = "arXiv",
    primaryClass = "hep-th",
    reportNumber = "CERN-TH-2019-143",
    doi = "10.1007/JHEP10(2019)085",
    journal = "JHEP",
    volume = "10",
    pages = "085",
    year = "2019"
}

@article{Gromov:2019aku,
    author = "Gromov, Nikolay and Sever, Amit",
    title = "{Derivation of the Holographic Dual of a Planar Conformal Field Theory in 4D}",
    eprint = "1903.10508",
    archivePrefix = "arXiv",
    primaryClass = "hep-th",
    reportNumber = "CERN-TH-2019-039",
    doi = "10.1103/PhysRevLett.123.081602",
    journal = "Phys. Rev. Lett.",
    volume = "123",
    number = "8",
    pages = "081602",
    year = "2019"
}

@article{Kade:2024lkc,
    author = "Kade, Moritz",
    title = "{The three-dimensional N=2 superfishnet theory}",
    eprint = "2410.18176",
    archivePrefix = "arXiv",
    primaryClass = "hep-th",
    reportNumber = "HU-EP-24/30-RTG",
    doi = "10.1007/JHEP01(2025)100",
    journal = "JHEP",
    volume = "01",
    pages = "100",
    year = "2025"
}

@article{Kazakov:2018qbr,
    author = "Kazakov, Vladimir and Olivucci, Enrico",
    title = "{Biscalar Integrable Conformal Field Theories in Any Dimension}",
    eprint = "1801.09844",
    archivePrefix = "arXiv",
    primaryClass = "hep-th",
    doi = "10.1103/PhysRevLett.121.131601",
    journal = "Phys. Rev. Lett.",
    volume = "121",
    number = "13",
    pages = "131601",
    year = "2018"
}

@article{BeisertIntegrability:2016,
    author = "Beisert, Niklas",
    title = {Introduction to Integrability},
    year = "2016",
    note = "lecture notes (2016)",
    url = "https://people.phys.ethz.ch/~nbeisert/lectures/"
}

\end{bibtex}

\bibliographystyle{nb}
\bibliography{\jobname}

\end{document}